\documentclass[11pt]{iopart}
\usepackage{iopams}
\usepackage{setstack}
\usepackage{hyperref} 
\usepackage{bm}
\usepackage{longtable}
\usepackage{graphicx}
\usepackage[margin=2cm]{geometry}
\usepackage{pdflscape}
\usepackage{cite}
\usepackage{subcaption}
\hypersetup{ colorlinks=true, linktoc=all, linkcolor=blue, }

\makeatletter
\newcommand\footnoteref[1]{\protected@xdef\@thefnmark{\ref{#1}}\@footnotemark}
\makeatother

\begin{document}

\title{Modelling the emergence of cosmic anisotropy from non-linear structures}

\author{Theodore Anton$^1$ and Timothy Clifton$^2$}
\address{Department of Physics \& Astronomy, Queen Mary University of London, UK.}
\ead{$^1$t.j.anton@qmul.ac.uk, $^2$t.clifton@qmul.ac.uk}

\begin{abstract}

Astronomical observations suggest that the Universe may be anisotropic on the largest scales. In order to model this situation, we develop a new approach to cosmology that allows for large-scale anisotropy to emerge from the growth of non-linear structure. This is achieved by decomposing all relevant fields with respect to a preferred space-like direction, and then averaging the resulting scalar quantities over spatial domains. Our approach allows us to derive a set of large-scale effective field equations that govern the dynamics of any emergent large-scale anisotropy, and which (up to back-reaction terms) take the form of the field equations of the locally rotationally symmetric Bianchi cosmologies. We apply our approach to the dust-filled Farnsworth solutions, which are an interesting set of exact cosmological models that allow for both anisotropic expansion and large-scale bulk flow.

\end{abstract}

\section{Introduction}\label{sec:introduction}

The standard paradigm for interpreting cosmological observations treats the Universe as being homogeneous and isotropic on large scales. This is justified in part by the near perfect isotropy of the CMB \cite{Planck_2018}, but has led to a cosmological model that appears to be in tension due to the non-vanishing of some direction-dependent observables \cite{Aluri_2022} as well as due to self-inconsistencies \cite{Hubble1, Hubble2, Hubble, s81, s82}. In particular, the values of cosmological parameters inferred from the CMB appear to show some directional dependence \cite{CMB1, CMB2}, an apparent hemi-spherical power asymmetry \cite{hemi1, hemi2} and the alignment of low-$l$ multipoles \cite{lowl1, lowl2}. In the more nearby Universe there are asymmetries in galaxy cluster scaling relationships \cite{cgs}, observations of radio galaxies \cite{radio}, quasars \cite{quasars} and perhaps even supernovae \cite{sne}. Beyond this, there is a long history of larger-than-expected bulk flows on large scales, with controversial results on a ``dark flow'' \cite{df} through to more recent results \cite{bulk}.

These results are certainly not without controversy, sometimes with regard to the specifics of the approach used when inferring them, and often with regard to the statistical significance of finding what appear to be anomalies in large and varied collections of data sets. We do not aim to argue for or against the veracity or significance of any of these results, but merely to note that if there exist data that suggest the Universe may not be isotropic on large scales, then it should also be of interest to know how large-scale anisotropy may or may not be able to emerge within an inhomogeneous universe that contains non-linear structure, but which itself started off close to homogeneous and isotropic at early times. This is our present purpose.

From a theoretical perspective, the possible emergence of anisotropy at late times is itself highly intriguing. It was suggested long ago by Ellis \& Baldwin \cite{Ellis_1984} that any misalignment between the frame in which the galaxy/quasar distribution is at rest, and the frame in which the dipole of the CMB vanishes, could be used as a test of large-scale isotropy. Indeed, the idea that observers may be ``tilted'' with respect to a preferred foliation of homogeneous spaces has itself recently been the subject of much study \cite{Tsagas_2009, Tsagas_2011, Tsagas_2021, Tsagas_2021_b, Santiago_2022}, with suggestions that the consequences for some observables might even be large enough to nullify the evidence for dark energy entirely \cite{Colin_2019, Mohayaee_2020, Mohayaee_2021}. These proposals have so far gone without a concrete mechanism by which such large-scale misalignments could be generated. We take this as further motivation for the development of a framework for studying emergent anisotropy, which could then be used to evaluate the possibility of these scenarios being realised.

An important aspect in our approach is the concept of ``emergence'', which is intended to describe the possibility of the properties in which we are interested being realised on average, from a space-time which is highly inhomogeneous on small scales. This concept is vital for the goal we have outlined so far, as anisotropy in exactly homogeneous cosmologies typically decays rapidly in time \cite{ani}, meaning that any anisotropy in the late universe would mean enormous anisotropy at early times. This is clearly not physically realistic, and so the only other possibility would be to have large-scale anisotropy realised as an emergent property of a more complicated inhomogeneous space-time. We expect that anisotropy generated in this way should weaken the tight bounds that could otherwise be imposed from the CMB \cite{Pontzen_2007, Pontzen_2009, Saadeh_2016}, and might even allow for the intruiging fits of Bianchi VII$_h$ templates to the observational data to be reinterpreted \cite{7h1, 7h2, 7h3, 7h4}.

The approach we will use in our construction will follow the general philosophy of the scalar averaging formalism developed by Buchert \cite{Buchert_2000}, which works by covariantly averaging locally-defined scalars over three-dimensional spatial volumes. This formalism is designed to include the consequences of the non-commutativity of averaging and evolution under Einstein's equations, and results in a set of effective field equations that govern the behaviour of large-scale averages. These equations take the same form as the locally defined versions that are more frequently used in the standard approach to cosmological modelling, but with ``back-reaction'' terms that account for the effects of non-linear structure on the cosmological dynamics. This is exactly what we require for our current purpose, although we will need to extend and generalize Buchert's approach in a number of regards in order to make it suitable for the task at hand.

First, if we are to model situations in which matter is to have a bulk motion then we will need to perform our averaging procedure on a set of hypersurfaces that are {\it not} necessarily orthogonal to the flow of matter \cite{Umeh_2011}. This will require generalizing the standard approach, in which the averaging surfaces are usually chosen such that the matter is comoving. Second, we will need to introduce a preferred spatial direction, and covariantly decompose all geometric, kinematic and matter field quantities with respect to it. This will allow us to average all resultant scalar quantities on our generalized foliation, and identify back-reaction terms by comparing to the field equations of the locally rotationally symmetric (LRS) Bianchi or Kantowski-Sachs cosmologies. Our many back-reaction terms will generalize the single term that occurs when considering only the isotropic part of the expansion, and will allow us to study if and how anisotropic cosmological expansion and bulk flows could emerge from inhomogeneous space-times.

This paper is arranged as follows: in Section \ref{sec:formalism} we introduce our mathematical formalism, then in Section \ref{sec:apply} we discuss how this formalism can be applied. The resulting equations are presented and discussed in Section \ref{sec:averaged_equations}, before being applied to an illustrative example in Section \ref{sec:farnsworth}. Finally, we offer some concluding remarks in Section \ref{sec:discussion}. We choose units such that $c = 8\pi G = 1$ throughout, and use metric signature $\left(-,+,+,+\right)$. Bold typeface is used to denote vectors.

\section{Mathematical formalism}\label{sec:formalism}

Let us now introduce the mathematical formalism that will be necessary to identify all of the required kinematic, curvature and matter quantities, to relate these quantities to those that would be measured by a set of observers, and to construct suitable, well-defined spatial averages of those local quantities, from which our concept of an emergent cosmological model can be realised. 

\subsection{The 1+1+2 covariant decomposition}\label{subsec:112}

In order to describe a cosmology with a preferred space-like direction, such as the symmetry axis of an LRS model, it is useful to perform a semi-tetrad decomposition of the field equations with respect to both a preferred time-like vector $\mathbf{n}$ and a preferred space-like vector $\mathbf{m}$ (satisfying $m_a n^a = 0$, $n_a n^a = -1$, and $m_a m^a = 1$). One can then decompose all vector and tensor quantities with respect to $\mathbf{n}$ and $\mathbf{m}$, while computing the Ricci and Bianchi identities for $\mathbf{n}$ and $\mathbf{m}$ gives rise to a set of equations which is physically equivalent to Einstein's equations. This $1+1+2$ formulation was first introduced by Greenberg \cite{Greenberg_1970}, developed in Refs. \cite{Tsamparlis_1983, Mason_1985, Zafiris_1997, Clarkson_2003, Betschart_2004, Burston_2006_a, Burston_2006_b}, and worked out in full by Clarkson \cite{Clarkson_2007} and Keresztes {\it et al.} \cite{Keresztes_2015}. We will introduce the salient features of this approach here, and explain why they are of use for our approach.

Let us suppose that we have a time-like vector $\mathbf{n}$. One can then define the projection tensor $f_{ab} = g_{ab} + n_a n_b$, as well as derivative operators along and orthogonal to {\bf n}:
\begin{equation}
\dot{\phantom{a}} \equiv n^a \nabla_a \qquad {\rm and} \qquad 
D_a T_{b c} \equiv f_a^{\phantom{a} d}  f_b^{\phantom{b} e}  f_c^{\phantom{c} f} \nabla_d T_{ef} \, ,
\end{equation}
with obvious generalizations to tensors of other ranks. These allow us to decompose the covariant derivative of $\mathbf{n}$ into kinematic quantities as \cite{Ellis_1999}
\begin{equation}
\nabla_a n_b = -n_a \dot{n}_b + \frac{1}{3}\Theta f_{ab} + \sigma_{ab} + \eta_{cab}\omega^c,
\end{equation}
where $\dot{n}_a = n^b \nabla_b n_a$, $\Theta = D_a n^a$, $\sigma_{ab}= D_{\langle a}n_{b\rangle}$ and $\omega_a = \frac{1}{2}\eta_{abc}D^b n^c$. These are, respectively, the acceleration of observers with 4-velocity $\mathbf{n}$, the isotropic expansion of the space-like 3-surfaces orthogonal to $\mathbf{n}$, and the volume-preserving shear and vorticity. In writing these expressions down, we have introduced angled brackets around indices to denote a tensor that is projected orthogonally to $\mathbf{n}$ and made symmetric and trace-free, such that e.g.
\begin{equation}
\sigma_{ab} = D_{\langle a}n_{b\rangle} \equiv \left(f_{(a}^{\ c} f_{b)}^{\ d} - \frac{1}{3}f_{ab}f^{cd}\right)\nabla_c n_d \, .
\end{equation}
We have also used $\eta_{abc} \equiv n^d \eta_{dabc}$ as the alternating tensor on the 3-surfaces orthogonal to $\mathbf{n}$.

Let us now further suppose that we have a space-like vector $\mathbf{m}$, which is orthogonal to $\mathbf{n}$. The covariant derivative of $\mathbf{m}$ can be decomposed as
\begin{equation}\label{nabla_a m_b}
\nabla_a m_b = - n_a \dot{m}_b + D_a m_b  + n_b m^c D_a n_c + n_a n_b m^c \dot{n}_c \, ,
\end{equation}
and a projection tensor onto the two-dimensional ``screen spaces'' orthogonal to both $\mathbf{n}$ and $\mathbf{m}$ can be defined as
\begin{equation} \label{bigm}
M_{ab} = f_{ab} - m_a m_b = g_{ab} + n_a n_b - m_a m_b \, .
\end{equation}
It can be seen that the projected time and space derivatives of $\mathbf{m}$ can be written as
\begin{equation}\label{D_a m_b}
\dot{m}_a = \mathcal{A} \, n_a + \alpha_a \qquad {\rm and} \qquad D_a m_b = m_a a_b + \frac{1}{2}\phi M_{ab} + \xi\epsilon_{ab} + \zeta_{ab}\, ,
\end{equation}
where $\epsilon_{ab} = \eta_{abc}m^c$ is the alternating tensor on the screen space. The new quantities introduced in Eq. (\ref{D_a m_b}) are $\alpha_a = M_{ac} n^b \nabla_b m^c$ and $\mathcal{A} =  -n^a \dot{m}_a =m_a \dot{n}^a$, which are the projections of $\dot{m}_a$ orthogonal and parallel to ${\bf n}$, and $a_b = m^c D_c m_b$, $\phi = M^{ab} D_a m_b$,
$\zeta_{ab} = \left[M^{\: c}_{(a}M^{\: d}_{b)} - \frac{1}{2}M_{ab}M^{cd}\right]D_{\langle c} m_{d \rangle}$, and $\xi = \frac{1}{2}\epsilon_{ab}M^{ac}M^{bd} D_c m_d$, which are the non-geodesy of $\mathbf{m}$, and the expansion, shear and twist of the screen space along {\bf m}. We use the names ``expansion'' and ``shear''  at this point in analogy with the quantites defined for {\bf n}, despite $\mathbf{m}$ being a space-like vector. This is done for linguistic ease.

To go further, we note that all 3-vectors and 3-tensors defined with respect to {\bf n} can be further projected parallel and orthogonal to {\bf m}. For 3-vectors $v_a = v_{\langle a \rangle} = f_a^{\ b}v_b$, this can be written as
\begin{equation} \label{vdef}
v_a = V m_a + V_a \, ,
\end{equation}
where $V = v_b m^b$ is the projection of $v_a$ parallel to $m_a$, and $V_a = M_a^{\ b}v_b$ is the projection of $v_a$ into the screen spaces. Similarly, for projected, symmetric and tracefree 3-tensors $t_{ab} = t_{\langle ab\rangle}$, we can write
\begin{equation} \label{tdef}
t_{ab}= \tau\left(m_a m_b - \frac{1}{2}M_{ab}\right) + 2 \tau_{(a}m_{b)} + \tau_{ab} \, ,
\end{equation}
where $\tau = t_{ab} m^a m^b$, $\tau_a = M_{ab}m_c t^{bc}$ and $\tau_{ab} = [M^{\: c}_{(a}M^{\: d}_{b)} - \frac{1}{2}M_{ab}M^{cd}]t_{cd}$
are the projections of $t_{ab}$ twice parallel, once parallel and once orthogonal, and twice orthogonal to ${\bf m}$, respectively.

Using this approach, we can decompose the acceleration and vorticity 3-vectors associated with {\bf n} in the following way:
\begin{eqnarray}
&& \dot{n}_a = \mathcal{A}m_a + \mathcal{A}_a \qquad {\rm and} \qquad \omega_a = \Omega m_a + \Omega_a \, .
\end{eqnarray}
The only other kinematic quantity is the shear of {\bf n}, which can be decomposed as
\begin{eqnarray}
\sigma_{ab} = \Sigma \left(m_a m_b - \frac{1}{2}M_{ab}\right) + 2\Sigma_{(a}m_{b)} + \Sigma_{ab} \, .
\end{eqnarray}
The quantities $\mathcal{A}$, $\Omega$ and $\Sigma$ are covariantly defined scalars, while $\mathcal{A}_a$, $\Omega_a$, $\Sigma_a$ and $\Sigma_{ab}$ are covariantly defined 2-vectors and tensors defined in the screen spaces orthogonal to {\bf n} and {\bf m}.

We have so far considered the decompositions of the derivatives of {\bf n} and {\bf m}. For a complete system we will also require quantities associated with the Ricci and Weyl curvature of the space-time. The former of these can be related to the matter content of the space-time using Einstein's equations, and therefore written in terms of a stress-energy tensor that can be decomposed with respect to {\bf n} as
\begin{equation}
T_{ab} = \mu n_a n_b + p f_{ab} + 2 q_{(a}n_{b)} + \pi_{ab},
\end{equation}
where $\mu = T_{ab}n^a n^b$, $p = \frac{1}{3}T_{ab}f^{ab}$, $q_a = -T_{bc}n^b f_a^{\ c}$, and $\pi_{ab} = T_{\langle ab \rangle}$ are the energy density, isotropic pressure, momentum density and anisotropic stress, respectively. The last two of these can be further decomposed with respect to {\bf m} to get
\begin{eqnarray}
&& q_a = Q m_a + Q_a \qquad
{\rm and} \quad \pi_{ab} = \Pi \left(m_a m_b - \frac{1}{2}M_{ab}\right) + 2\Pi_{(a}m_{b)} + \Pi_{ab} \, ,
\end{eqnarray}
where all quantities are defined as in Eqs. (\ref{vdef}) and (\ref{tdef}), above. Finally, the Weyl tensor can be decomposed with respect to {\bf n} into an ``electric'' and a ``magnetic'' part, as
\begin{equation}
E_{ab} = C_{acbd}n^c n^d \qquad {\rm and} \qquad H_{ab} =  \frac{1}{2}\eta_{efda}C^{ef}_{\quad bc} n^c n^d \, ,
\end{equation}
which can be further decomposed with respect to the space-like vector {\bf m} as
\begin{eqnarray*}
\hspace{-2cm}E_{ab} = \mathcal{E} \left(m_a m_b - \frac{1}{2}M_{ab}\right) + 2\mathcal{E}_{(a}m_{b)} + \mathcal{E}_{ab}, 
\qquad H_{ab} = \mathcal{H} \left(m_a m_b - \frac{1}{2}M_{ab}\right) + 2\mathcal{H}_{(a}m_{b)} + \mathcal{H}_{ab}\, .
\end{eqnarray*}
Again, all quantities in these expressions are defined as in Eqs. (\ref{vdef}) and (\ref{tdef}). This completes the decomposition of all required quantities with respect to both {\bf n} and {\bf m}.

The full and irreducible set of quantities describing the kinematics of {\bf n} and {\bf m}, the matter variables and the Weyl curvature is therefore given as follows:

\begin{equation}\label{112scalars}
\hspace{-1.7cm}{\rm 2\mbox{-}scalars:} \quad \left \lbrace \Theta, \mathcal{A}, \Sigma, \Omega, \phi, \xi, \mathcal{E}, \mathcal{H}, \mu, p, Q, \Pi \right \rbrace 
\end{equation}

\begin{equation}
\hspace{-1.7cm}{\rm 2\mbox{-}vectors:} \quad\left \lbrace \mathcal{A}_a, \Sigma_a, \Omega_a, \alpha_a, a_a, \mathcal{E}_a, \mathcal{H}_a, Q_a, \Pi_a \right\rbrace
\end{equation}

\begin{equation}
\hspace{-1.7cm}{\rm 2\mbox{-}tensors:} \quad\left \lbrace \Sigma_{ab}, \zeta_{ab}, \mathcal{E}_{ab}, \mathcal{H}_{ab}, \Pi_{ab} \right \rbrace
\end{equation}

\noindent
The reader is referred to Refs. \cite{Clarkson_2007} and \cite{Keresztes_2015} for the further details of these decompositions, and to \ref{sec:notation} for a full list of all quantities defined above. For the remainder of this paper we will take $\Omega = 0 =\Omega_a$, so that $\mathbf{n}$ is hypersurface-orthogonal. This is required for having a foliation of space-like 3-surfaces over which averaging can be performed. It is not a restriction on the properties of matter. 

\subsection{Changing foliation}\label{subsec:tilted_averaging}

We have so far not specified anything about the time-like vector {\bf n}, other than it being hypersurface-forming. In the sections that follow we will consider specific choices of this vector, which are orthogonal to foliations with certain special properties. Here, we will simply note that it will be of interest to be able to transform between foliations, and present some of the mathematics that is required to do so.

\begin{figure}[h]
\centering
\includegraphics[width=0.8\linewidth]{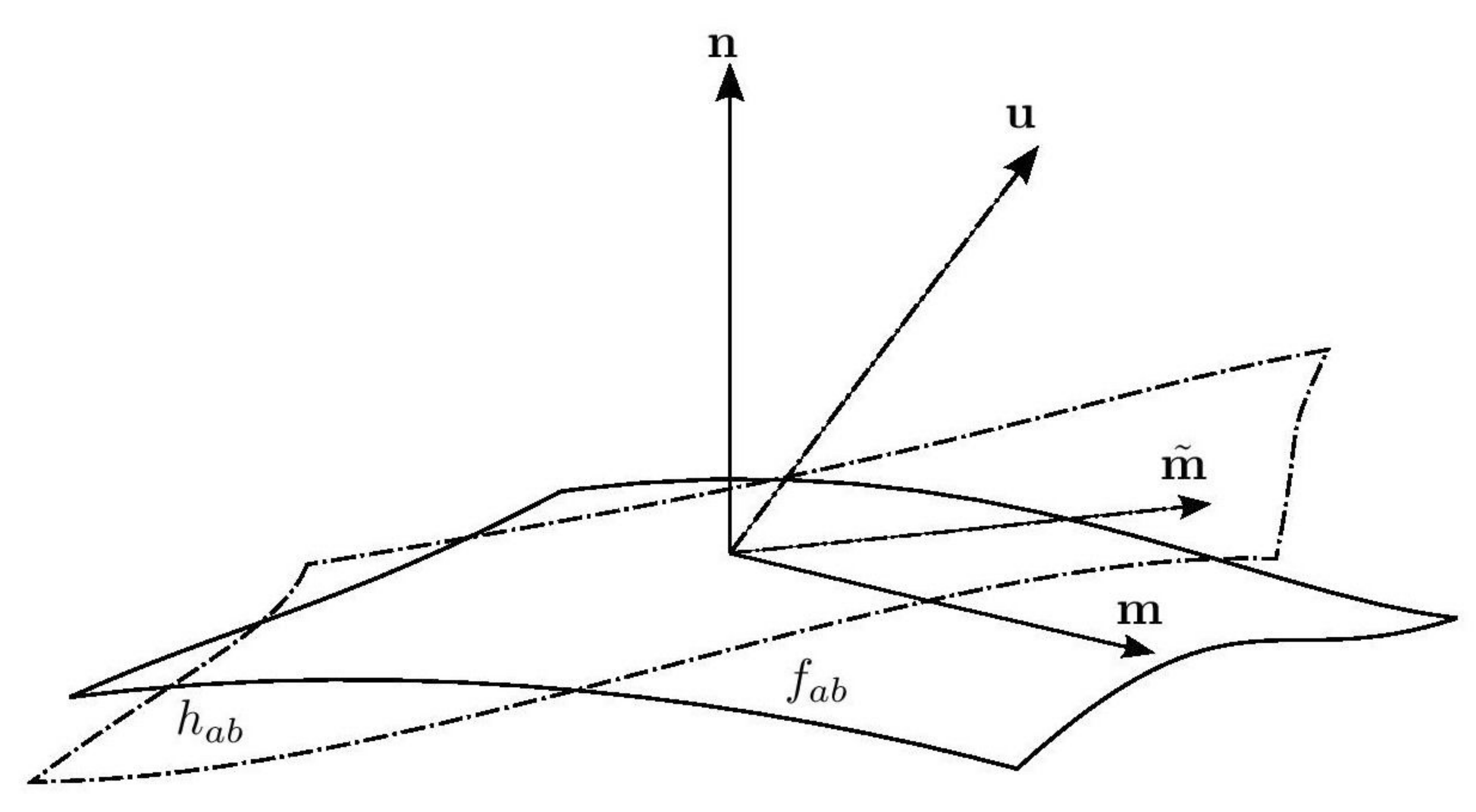}
\caption{Visualisation of foliations induced by the time-like vectors $\mathbf{n}$ and $\mathbf{u}$, with the associated projection tensors $f_{ab}$ and $h_{ab}$, and the preferred space-like vectors $\mathbf{m}$ and $\tilde{\mathbf{m}}$.}
\label{fig1}
\end{figure}

In order to understand this, let us consider a boost from {\bf n} to some second time-like vector {\bf u}. The components of these vectors are related by
\begin{equation}\label{boost}
u_a = \gamma\left(n_a - v_a\right) \qquad {\rm and} \qquad n_a = \gamma\left(u_a + w_a\right) \, , 
\end{equation}
where $v_a = \gamma^{-1}f_a^{\ b} w_b$ and $\gamma = \left(1-v^2\right)^{-1/2} = \left(1-w^2\right)^{-1/2}$. Here the boost vectors $v^a$ and $w^a$ exist respectively in the surfaces orthogonal to {\bf n} and {\bf u}, such that $n_a v^a = u_a w^a = 0$. The projection tensor into the hypersurfaces orthogonal to {\bf u} (if they exist) can then be defined as $h_{ab} = g_{ab} + u_a u_b$, which allows us to write $w_a = \gamma^{-1} h_a^{\ b} v_b$. The same projection tensor also allows us to define a new vector $\mathbf{\tilde{m}}$ with components
\begin{equation}\label{projectm}
\tilde{m}_a = k \, h_a^{\ b} m_b \, , \qquad {\rm such \;\; that} \qquad m_a = \tilde{k} \, f_a^{\ b} \tilde{m}_b \,,
\end{equation}
where $k = \left[1+\left(\gamma m_c v^c\right)^2\right]^{-1/2}$ and $\tilde{k} = \left[1+\left(\gamma \tilde{m}_c w^c\right)^2\right]^{-1/2}$. These are the components of the preferred space-like vector {\bf m} projected orthogonal to {\bf u}, as illustrated in Figure \ref{fig1}.

The whole system of $1+1+2$-equations from Section \ref{subsec:112} can now be set up with respect to either {\bf n} or {\bf u}, with the preferred space-like vector projected according to Eq. (\ref{projectm}). After we boost from $\mathbf{n}$ to $\mathbf{u}$, we wish to express the scalars in the new foliation in terms of those in the old one, plus any terms arising from the boost 3-velocity $\mathbf{v}$ that connects them. If $\left\vert \mathbf{v} \right\vert \ll 1$, which is expected if, for example, $\mathbf{n}$ corresponds to the canonical foliation of FLRW spacetime, and $\mathbf{u}$ is the matter 4-velocity in that spacetime, then one can expand these expressions order by order in $\mathbf{v}$, leading to vast simplifications. However, in the general case, one need not make this approximation. Performing the full calculations, one finds the following results:
\vspace{0.25cm}

The isotropic expansion $\tilde{\Theta} = h^{ab} \nabla_a u_b$ is related to $\Theta = f^{ab}\nabla_a n_b$ by 
\begin{equation}
\tilde{\Theta} = \gamma\left[\Theta - \kappa - v_a\left(\dot{n}^a + \gamma^2 \dot{v}^a\right) + \gamma^2\left(\frac{1}{3}\kappa v^2 + \beta_{ab} v^a v^b\right)\right],
\end{equation}
where $\kappa = D_a n^a$ and $\beta_{ab} = D_{\langle a} v_{b\rangle}$. This equation shows how the expansion of the rest spaces defined with respect either {\bf n} or {\bf u} are related to each other, and explicitly provides the additional terms that contribute to the expansion after performing a boost. 
\vspace{0.25cm}

The scalar shear $\tilde{\Sigma} = \left(h_a^{\ c} h_b^{\ d} - \frac{1}{3}h_{ab}h^{cd}\right)\nabla_c u_d \tilde{m}^a \tilde{m}^b$ is related to $\Sigma$ by
\begin{eqnarray}
\hspace{-1.5cm}
\tilde{\Sigma} &=& k^2\gamma \Sigma + \frac{\gamma}{3}\left(\Theta-\kappa\right)\left(k^2\gamma - 1\right) + \frac{\gamma}{3}v_a\left(\dot{n}^a+\gamma^2\dot{v}^a\right) - \frac{\gamma^3}{3}\left(\frac{\kappa}{3}v^2 + \beta_{ab}v^a v^b\right) \\
\nonumber \hspace{-2.5cm} && + k^2\gamma \Bigg[-\beta_{ab}m^am^b -\left(\gamma m_a v^a\right)^2 +\gamma^2 m^a v^b\left(m_c v^c - 1\right)\left(\beta_{ab}+W_{ab}\right) - \gamma^2 m_a v^a\mathcal{A} \\
\nonumber \hspace{-2.5cm }&&\hspace{2cm} + \gamma^2 m_b v^b\left(\frac{1}{3}\left(3\Sigma+\Theta-\kappa\right) m_a v^a + \Sigma_a v^a + \dot{v}_a \left(m^a + \gamma^2 m_c v^c v^a\right)\right) 
\\
\nonumber \hspace{-2.5cm }&&\hspace{8cm }
 - \gamma^4 \left(m_c v^c\right)^2\left(\frac{1}{3}\kappa v^2 + \beta_{ab} v^a v^b\right) \Bigg],
\end{eqnarray}
and the scalar acceleration, $\tilde{\mathcal{A}} = \tilde{m}^a u^b \nabla_a u_b$, and the scalar vorticity $\tilde{\Omega} = \frac{1}{2}\tilde{m}^a \eta_{dabc}u^d \nabla^b u^c$, are
\begin{equation}
\tilde{\Omega} = -k\gamma^2\left[W_a\left(m^a + \gamma^2 m_bv^b v^a\right) - \frac{1}{2}\gamma^2 m_a v^a \chi_{bc}W^{bc}\right],
\end{equation}
and
\begin{equation}
\hspace{-1.5cm}
\tilde{\mathcal{A}} = k\gamma^2\left[\mathcal{A} - m_a \dot{v}^a - \gamma^2 m_a v^a v_b \dot{v}^b - \frac{1}{3}m_a v^a\left(3\Sigma + \Theta - \kappa\right) - \Sigma_a v^a + v^a m^b\left(\beta_{ab} + W_{ab}\right)\right]  ,
\end{equation}
where we have used that $\Omega = \frac{1}{2} m^a \eta_{dabc}n^d \nabla^b n^c = 0$. One sees immediately that if the vorticity $W_a$ associated with the relative velocity $\mathbf{v}$ vanishes, then $\tilde{\Omega} = 0$ as well.
\vspace{0.25cm}

Meanwhile, the twist and expansion of the screen spaces along $\tilde{\mathbf{m}}$, $\tilde{\xi} = \frac{1}{2}\tilde{\epsilon}_{ab} \tilde{M}^{ac}\tilde{M}^{bd}\tilde{D}_c \tilde{m}_d$ and $\tilde{\phi} = \tilde{M}^{ab}\tilde{D}_a \tilde{m}_b$, are related to $\xi$ and $\phi$ by
\begin{equation}
\hspace{-1.5cm}
\tilde{\xi} = k^2\gamma\left[\xi + \epsilon_{ab}v^a\left(\alpha^b + \Sigma^b\right)\right].
\end{equation}
and
\begin{eqnarray}
\hspace{-1.5cm}
\tilde{\phi} &=& k\phi + k\mathcal{A} - \tilde{A} - k\gamma m_a v^a \tilde{\Theta} + \left(m^a - \gamma^2 m_b v^b\left(n^a - v^a\right)\right)\nabla_a k \\
\hspace{-2.5cm} \nonumber && + k\gamma^2 \left[-\alpha_a v^a + m_a v^a\left(a_b v^b + \frac{\kappa}{3}\right) - m_a\dot{v}^a + \left(\frac{\phi}{2} M_{ab} + \zeta_{ab}\right)v^a v^b  + \left(\beta_{ab} + W_{ab}\right) v^a m^b \right] \\
\hspace{-2.5cm} \nonumber &&+ k\gamma^4 m_c v^c\left(-v_a\dot{v}^a + \frac{\kappa}{3}v^2 + \beta_{ab}v^a v^b\right),
\end{eqnarray}
\begin{eqnarray}
\hspace{-2.5cm}
\nonumber {\rm where} \quad \nabla_a k &=& - \gamma^2 k^3 m_c v^c \Bigg[\gamma^2 m_b v^b v^d\left(-n_a \dot{v}_d + \beta_{ad} + W_{ad}\right) + \frac{\gamma^2}{3}m_b v^b \kappa v_a - \alpha_b v^b n_a \\
\nonumber \hspace{-2.5cm} && + v^b\left(\frac{\phi}{2}M_{ab} + \zeta_{ab} + \xi \epsilon_{ab} + m_a a_b\right) - \dot{v}_b m^b n_a + \frac{\kappa}{3}m_a + m^b\left(\beta_{ab} + W_{ab}\right)\Bigg].
\end{eqnarray}

For the matter variables, the energy density $\tilde{\mu} = T_{ab}u^a u^b$, isotropic pressure $\tilde{p} = \frac{1}{3}T_{ab}h^{ab}$, and scalar heat flux $\tilde{Q} = -T_{bc}u^b h^c_{\ a} \tilde{m}^a$ are given by
\begin{eqnarray}
\tilde{\mu} &=& \gamma^2\left(\mu + pv^2 + 2q_a v^a + \pi_{ab}v^av^b\right) , \\
\tilde{p} &=& \left(1+\frac{1}{3}\gamma^2 v^2\right) p + \frac{1}{3}\gamma^2 v^2 \mu + \frac{1}{3}\gamma^2\left(2q_a v^a + \pi_{ab}v^a v^b\right) , \\
\tilde{Q} &=& k\gamma\left[Q + m_c v^c\left(\gamma^2\left(\mu+p+2q_a v^a + \pi_{ab}v^a v^b\right) + \Pi\right) + \Pi_a v^a \right] ,
\end{eqnarray}
and the scalar anisotropic stress $\tilde{\Pi} = \left(h_a^{\ c} h_b^{\ d} - \frac{1}{3}h_{ab}h^{cd}\right)T_{cd} \tilde{m}^a \tilde{m}^b$ is given by
\begin{eqnarray}
\hspace{-1.5cm}
\tilde{\Pi} &=& \gamma^2 \mu\left(\left(k\gamma m_a v^a\right)^2 - \frac{1}{3}v^2\right) + p \left[ k^2 - 1 + 2\left(k\gamma m_a v^a\right)^2   + \frac{1}{3}\gamma^2 v^2\left(3\left(k\gamma m_a v^a\right)^2 - 1\right)\right] \\
\hspace{-2.5cm} \nonumber &&+ 2k^2\gamma^2 m_a v^a Q + k^2 \Pi\left(1+2\left(\gamma m_a v^a\right)^2\right)  + \frac{\gamma^2}{3}\left(3\left(k\gamma m_c v^c\right)^2-1\right)\left[2q_a v^a + \pi_{ab} v^a v^b\right].
\end{eqnarray}

Finally, one can compute the scalar parts of the Weyl curvature for the new congruence. For the scalar parts of the electric and magnetic Weyl tensors this gives \cite{Ellis_2010} 
\begin{eqnarray}
\tilde{\mathcal{E}} &=& k^2\gamma^2\left[\left(1+v^2 - \left(m_a v^a\right)^2 - \frac{1}{2}M_{ab}v^a v^b\right)\mathcal{E} + 2\chi^{ab}v_c m_a\mathcal{H}_b + \mathcal{E}_{ab}v^a v^b\right] , \\
\tilde{\mathcal{H}} &=& k^2\gamma^2\left[\left(1+v^2 - \left(m_a v^a\right)^2 - \frac{1}{2}M_{ab}v^a v^b\right)\mathcal{H} - 2\chi^{ab}v_c m_a\mathcal{E}_b + \mathcal{H}_{ab}v^a v^b\right].
\end{eqnarray}
This completes the transformation rules for all of our covariantly defined scalars. The 2-vector and tensor quantities that result from the $1+1+2$ decomposition can also be transformed between foliations, but only appear in our averaged description via ``backreaction'' terms. We therefore omit presenting their transformation rules here. 

In what follows, we will use often use {\bf u} to refer to the flow lines of an irrotational dust fluid. In the presence of such matter, this vector is uniquely defined and therefore also provides a unique foliation on which to perform averaging. This is the foliation that is most often used in the literature on this subject, and is the one chosen for the standard approach to formulating Buchert's equations \cite{Buchert_2000}. We will then use {\bf n} to refer to any other foliation, chosen (for example) for observational or geometric reasons. Such freedom is required if we are to allow for the possibility of bulk flows, as the 3-velocity of matter vanishes identically if we choose the foliation to be induced by the flow of dust {\bf u}. It also allows us the freedom to re-foliate in situations in which the description of space-time might break down if it were specified by the fluid flow, as would happen, for example, in perturbed FLRW cosmologies that contain non-linear structures, or if one wished to consider fluids with non-zero vorticity or caustics in their flow lines.

\subsection{Scalar averaging}\label{subsec:averaging}

Averaging in general relativity is a notoriously difficult problem to define, primarily as the theory is defined in terms of tensors, which cannot be straightforwardly compared at different points in a curved space-time \cite{vdh}. This makes defining averages difficult, as quantities that exist at different points in an averaging domain first have to be transported to the same point for comparison, which is a non-unique and path-dependent process. The problem is further exacerbated by the non-linearity of Einstein's equations, which means that the acts of averaging and evolution do not commute. The evolution of an averaged quantity cannot therefore be guaranteed to be the same as an average taken at some future time.

The first of these problems has been the focus of a large number of different studies, using a variety of different approaches (see e.g. Refs. \cite{Green_2011, Green_2013, Green_2014, Mars_1997, Zalaletdinov_1997, Coley_2005}). The benefits and drawbacks of each of these approaches are a subject of much debate, but in order to make progress one can consider a more limited case: the averaging of scalars. Such quantities can be readily compared at different points in averaging domains, and it is therefore much more straightforward to compute their averages. Indeed, attempts have been made to construct averaging schemes that completely characterize space-times {\it only} in terms of the averages of scalars \cite{Coley_2010}. In what follows we will proceed by constructing an averaging formalism for the scalars that result from our 1+1+2 decomposition.

We will be interested in spatial averages of scalar quantities, over the 3-dimensional spatial volumes orthogonal to the irrotational flow lines {\bf n}. Following Buchert \cite{Buchert_2000}, we will define these as follows:
\begin{equation} \label{avop}
\left\langle S \right \rangle := \frac{1}{V_{\mathcal{D}}}\int_{\mathcal{D}} \mathrm{d}V \: S \: = \: \frac{\int_{\mathcal{D}}\mathrm{d}^3 x\,\sqrt{f} \, S}{\int_{\mathcal{D}}\mathrm{d}^3 x\,\sqrt{f}} \, ,
\end{equation}
where $S$ is the scalar being averaged over the spatial domain $\mathcal{D}$, which has volume $V_{\mathcal{D}}=\int_{\mathcal{D}}\mathrm{d} V$, and where $\mathrm{d} V =\mathrm{d}^3 x\,\sqrt{f}$ is the spatial volume element. In these expressions, we have taken $x^i$ to be coordinates on the spatial hypersurface orthogonal to {\bf n}, and written $f= {\rm det}(f_{ab})$ where $f_{ab}=g_{ab}+n_an_b$ is the induced metric. This definition of averaging depends on our choice of foliation and on the selected averaging domain, but is otherwise unambiguous.

The rate of change of this average can be determined by considering that the factor $\sqrt{f}$ represents the volume of an infinitessimal region of space. Assuming the spatial coordinates are fixed by the flow of {\bf n}, we then have $(\sqrt{f})^{\boldsymbol{\cdot}} = \Theta \sqrt{f}$, where the dot represents the time derivative operator $n^a \nabla_a$ and where $\Theta = \nabla_a n^a$ is the expansion scalar. This allows one to derive Buchert's commutation rule for time derivative and averaging operations \cite{Buchert_2000}:
\begin{equation} \label{comop}
\left\langle S \right\rangle^{\boldsymbol{\cdot}} - \langle \dot{S}\rangle  = \left\langle S \Theta\right\rangle- \left\langle S \right\rangle\left\langle\Theta\right\rangle = {\rm Cov}(\Theta,S) \, ,
\end{equation}
where we have defined the covariance as ${\rm Cov}(S_1, S_2) :=  \left\langle S_1 S_2 \right\rangle - \left\langle S_1\right\rangle \left\langle S_2\right\rangle$. This result relies on the averaging domain $\mathcal{D}$ being propagated by the flow of {\bf n}.  In what follows we will make repeated use of the covariance of scalar quantities, as well as the variance, ${\rm Var}(S) := {\rm Cov}(S,S)$.

In Eq. (\ref{comop}) it is assumed that the lapse function $N$, between the leaves of the chosen foliation, is independent of spatial position. This is the case if (for example) the time-like 4-vector {\bf n} is identified with the 4-velocity of a field of irrotational dust {\bf u}, as in this case $\dot{n}^{a} = \dot{u}^a = 0$. Each leaf in the foliation can then be identified with a single value of the proper time along the world-lines of this dust, and will be the same at all points in each spatial hypersurface. More generally, we will have $\dot{n}_a = D_a \ln N \neq 0$, meaning that the proper time of a family of observers following {\bf n} is not constant across $\mathcal{D}$. For the purposes of calculating cosmological averages it is desirable to have quantities defined over an averaging domain to have a single value for the time coordinate $t$. This can be achieved by writing $n_a = - N \partial_a t$ where $N= \dot{t}^{-1}$, such that a derivative with respect to $t$ can be written as $\partial_t S = N \dot{S}$. The commutation rule (\ref{comop}) then becomes
\begin{equation}\label{commutation_lapseshift}
\partial_t\left\langle S\right\rangle - \left\langle \partial_t S\right\rangle = {\rm Cov}\left(N\Theta, S\right),   
\end{equation}
which allows us to calculate the evolution of scalar averages as a function of the coordinate time $t$, once the lapse function $N$ has been determined from the acceleration of {\bf n}.


One can apply the averaging operation from Eq. (\ref{avop}) to each of the terms in any scalar equation that is defined locally at each point in space-time. After suitable application of the commutation rule from Eq. (\ref{comop}) or (\ref{commutation_lapseshift}), one can then obtain from that averaged scalar equation a new one that is written entirely in terms averages and their derivatives with respect to time. Equations derived in this way can be thought of as governing the behaviour of averages, rather than the behaviour of locally-defined quantities (as usually occurs in theories of gravity). This is ideal for describing the emergent large-scale properties of a space-time, as is relevant for the study of cosmology. Indeed, the averaged equations that result from this procedure can often be written in a form identical to that of the local gravitational equations of an homogeneous cosmology, with any extra terms that occur being considered to describe the ``back-reaction'' effect that results from averaging small-scale structures.

In the standard approach to this problem, three scalar equations are usually considered: the Raychaudhuri equation for the evolution of the expansion, the Hamiltonian constraint equation that relates expansion and energy density, and a contracted Bianchi identity that corresponds to energy conservation. The averages of these are all evaluated in the foliation orthogonal to the flow of dust, and a set of Friedmann-like equations emerge that govern the average expansion \cite{Buchert_2000}. If one wishes to consider the emergence of any anisotropic quantity, however, then this necessarily requires further scalar equations, such as those governing the scalar shear $\Sigma$ and momentum density $Q$. This is what we will pursue in the following sections. We note that an interesting step in this direction has already been taken by extending the Buchert formalism to include an evolution equation for the magnitude of the shear, $\sigma^2 = \frac{1}{2} \sigma_{ab}\sigma^{ab}$ \cite{Barrow_2007}. We extend this result further by deriving the full set of averaged scalar equations in the $1+1+2$ formalism. We find a set of equations that can be used to interpret that large-scale cosmological average in terms of an emergent LRS Bianchi model with tilt, and with back-reaction terms that give the effect of small-scale structure on all emergent quantities.

\section{Application of the formalism}\label{sec:apply}

Let us now consider how to apply the formalism from Section \ref{sec:formalism} to a cosmological space-time. This will require a choice of the time-like and space-like vectors $\mathbf{n}$ and $\mathbf{m}$, and some consideration of how to understand and interpret the quantities that result.

\subsection{Choice of foliation}\label{subsec:fol_choice}

An application of the averaging procedure outlined above requires a choice of foliation, or equivalently a choice of the irrotational time-like 4-vector {\bf n}. The averages obtained using Eq. (\ref{avop}) will then correspond to the large-scale properties of the 3-dimensional spaces that constitute the leaves of this foliation. Different foliations will mean that one is considering different 3-dimensional spaces, and hence different averages will be obtained. It is therefore necessary to make sure an appropriate choice of foliation is made, for the situation being considered. This is given further importance by the fact that observers in different frames will infer different cosmological parameters from the Hubble flow around them \cite{Tsagas_2009, Tsagas_2011, Tsagas_2021, Tsagas_2021_b, Santiago_2022, Colin_2019, Mohayaee_2020, Mohayaee_2021}. 

In general, one might be interested in choosing a foliation that is expected to give results that can be associated with a particular observable \cite{syksyobs}, that have a particular mathematical or physical meaning associated with them \cite{cmc}, or that are perhaps convenient in some other way \cite{Umeh_2011}. For example, one may wish to construct Hubble diagrams in a frame comoving with the flow of matter \cite{Buchert_2000}, or in a frame of `most uniform Hubble flow' \cite{Wiltshire_2013, McKay_2015, Kraljic_2016}. Of course, in order to relate observables corresponding to quantities calculated on different foliations one will need to be able to transform between frames\footnote{We note that such transformations should not be confused with gauge dependence, as choice of foliation is in general a covariant and non-perturbative process, meaning that no background manifold is defined and that there is therefore no gauge issue \cite{Buchert_2012}.}, as considered in Section \ref{subsec:tilted_averaging}.

Let us now consider some specific choices of foliation that one might use:

\vspace{0.2cm}
\noindent
{\it The comoving foliation} exists when the Universe is filled with irrotational dust, and $\mathbf{n}$ is chosen to be coincident with the 4-velocity of that dust, $\mathbf{u}$. This choice shares some properties with the ``comoving synchronous'' gauge that may be familiar from studies of perturbed Friedmann-Lema\^{i}tre-Robertson-Walker (FLRW) models. It has the distinct benefit of being tied to a physical quantity: the flow of matter. It therefore takes a particularly privileged position in the pantheon of possible choices, but does require the matter flow to be vorticity-free, which is not expected to hold for realistic astrophysical structures. It also cannot account for the existence of any bulk flow, as it corresponds to the choice of frame in which matter is at rest. Nevertheless, this is the standard choice of foliation in much of the literature on mathematical cosmology \cite{Ellis_1999}, as well as in many studies of perturbed FLRW.

\vspace{0.2cm}
\noindent
{\it The gravitational rest frame} was defined in Ref. \cite{Umeh_2011} to be that in which the magnetic part of the Weyl tensor vanishes:
$$
H_{ab}=0 \,,
$$
so that the emergent cosmological model is as Newtonian in character as possible \cite{Maartens_1998}. Cosmologies that obey this condition have been described as ``silent'' \cite{silent}. Up to second order in perturbations around a Robertson-Walker geometry this choice corresponds to the Poisson gauge \cite{Umeh_2011}, but when applied non-perturbatively ends up being a strong restriction on the allowed solutions \cite{barnes, van_Elst_1997}. It therefore appears to be particularly useful for studies of perturbed FLRW models, as the Poisson gauge is one of the few that is expected to be valid into the regime of non-linear structure formation, where post-Newtonian expansions are required to describe weak gravitational fields \cite{poisson}.

\vspace{0.2cm}
\noindent
{\it The gravitational wave frame} is a more conservative choice defined by the demand that $H_{ab}$ is divergence-free, such that
$$
D^b H_{ab} = 0 \, .
$$
This is a covariant way of stating that the only gravito-magnetic contributions to the curvature measured by an observer comoving with  $\mathbf{n}$ would be those coming from gravitational waves \cite{hawking, Dunsby_1997}. It is therefore less restrictive than the condition $H_{ab} = 0$, and typically corresponds to a congruence which is distinct from the flow of dust \cite{Sopuerta_1999}. It is well-suited to numerical-relativistic cosmological simulations \cite{Heinesen_2022}, but in a general space-time it is not guaranteed that the frames it picks out are hypersurface forming (though one may be able to take the irrotational part, which would be).

\vspace{0.2cm}
\noindent
{\it The constant mean curvature foliation} has a long history \cite{cmc}, and is defined by the condition that the spatial gradient of the expansion scalar (also known as the ``mean curvature'') vanishes:
$$
D_a \Theta = 0\, .
$$
This foliation has the property of being unique (under certain circumstances), and having a monotonic variation in the expansion scalar between leaves \cite{rencmc}. It therefore provides a plausible candidate for the provision of a universal, cosmological arrow of time. The literature on this particular choice of foliation has been largely resticted to the literature on mathematical cosmology, but it also exists in perturbed FLRW models under the name of ``uniform expansion'' gauge.

\vspace{0.2cm}
The list of choices presented above is by no means exhaustive. Other interesting choices of foliation are specified by the ``zero-shear'' condition, $\sigma_{ab} =0$. This is closely related to the gravitational rest frame, in that the Ricci identities provide a constraint that sets $H_{ab}$ equal to the covariant curl of the shear tensor. Vanishing of the shear therefore corresponds to the vanishing of $H_{ab}$, which explains why the Poisson gauge (sometimes referred to as the ``zero shear'' gauge) corresponds to the gravitational rest frame. Another is the ``constant density'' foliation, in which the spatial gradient of the energy density vanishes, $D_a \mu = 0$. As with the comoving foliation, this has the benefit of being directly tied to the matter content of the space-time. However, it also shares the weakness that it will be highly problematic in the presence of non-linear structures.

\subsection{Choice of preferred spatial direction}

In order to extract scalars from anisotropic quantities, we also need to choose a preferred space-like direction $\mathbf{m}$. All 3-vector and tensor quantities can then be decomposed with respect to this direction as prescribed by Eqs. (\ref{vdef})-(\ref{tdef}), and the scalar parts of the decomposed quantities can be averaged according to Eq. (\ref{avop}). Clearly, in order to make sense, this vector must (in some sense) point in the same direction at every point in an averaging domain. Just as in the case of the time-like vector {\bf n} this presents a choice; there is in general not going to be any single uniquely preferred space-like direction {\bf m}, and we will need to select a suitable way to define this direction based on the situation at hand. This choice will be important for the outcome of our averaging process.

As in the case of choosing {\bf n}, we may wish to apply geometric or observational considerations when selecting a preferred direction {\bf m}. In space-times with symmetry, it may be that a preferred direction is selected by the existence of some Killing vectors. For example, in LRS geometries $\mathbf{m}$ can be unambiguously taken to correspond to the spatial axis around which the rotational symmetry exists. This can be identified by searching for the non-degenerate eigenvector of any non-vanishing 3-tensor, such as $\sigma_{ab}$ or $E_{ab}$ \cite{Clarkson_2007, vanElst_1996}. Alternatively, in algebraically special geometries it may be possible to pick out a unique direction from the projections of the canonical null tetrad into our chosen foliation. These will be specified by the properties of the Weyl tensor.

An alternative method for selecting a preferred spatial direction would be to use observational methods at each point within an averaging domain. This method would be better adapted to situations in which anisotropy in a particular observable is being considered, or where the space-time has no explicit symmetries or special algebraic properties (as is the case for the real Universe). For example, one might choose to take the axis of CMB parity asymmetry \cite{Zhao_2014, Cheng_2016}, the direction of greatest asymmetry in the galaxy distribution \cite{cgs, radio}, the quasar or Type Ia supernovae dipole \cite{quasars, sne}, or the direction of greatest variation of a cosmological parameter or coupling constant such as $H_0$ or the fine structure constant $\alpha$ \cite{Webb_2011, King_2012, Yeung_2022}. As long as these directions line up at different points in space, as one would expect in an anisotropic universe, then they should provide a well-motivated choice of preferred direction.

\subsection{Interpretation as an LRS Bianchi cosmology} \label{subsec:LRS}

As well as considering the choice of foliation and space-like direction, it seems worthwhile to also consider what the model that emerges from averaging the scalars in our 1+1+2-decomposition will represent. These will be precisely the set of scalars represented in Eq. (\ref{112scalars}). It is an extended set compared to the scalars that result from a 1+3-decomposition, and include (among other things) the scalar parts of 3-vectors such as the momentum density $Q$, and the scalar part of 3-tensors such as the shear $\Sigma$.

The averaged scalars that result from the 1+3-decomposition, as provided in Buchert's ground-breaking approach \cite{Buchert_2000}, are relatively straightforward to interpret. They result in a set of equations that govern the averages of the expansion scalar, the energy density of dust, and the curvature scalar of the 3-spaces:
$$
\{ \langle \theta \rangle, \langle \mu \rangle, \langle \hspace{-0.2cm}{\phantom{a}}^{(3)} \hspace{-0.1cm}R \rangle \} \, ,
$$
with everything else being collected together into ``back-reaction'' terms. Defining an effective scalar factor $a_{\mathcal{D}}$ and an effective curvature parameter $k_{\mathcal{D}}$ of the averaging domain $\mathcal{D}$ by
$$
\langle \theta \rangle = \frac{\dot{a}_{\mathcal{D}}}{a_{\mathcal{D}}} \qquad {\rm and } \qquad k_{\mathcal{D}} = \frac{\langle \hspace{-0.2cm}{\phantom{a}}^{(3)} \hspace{-0.1cm}R \rangle}{6 a_{\mathcal{D}}^2} \, ,
$$
the equations that result bear a striking similarity to the dust-dominated Friedmann equations that govern the behaviour of homogeneous and isotropic cosmologies \cite{Buchert_2000}. This is no accident, as the set of averaged scalars listed above are exactly those required to fully characterise an homogeneous and isotropic universe. All other quantities will be 3-vectors or 3-tensors, but these must vanish in geometries that are isotropic around every point in space. So, after discarding such quantities in the scalar averaging approach, we are left with exactly the set that is required to describe an FLRW model. One could say that the geometry has been averaged to a Friedmann cosmology.

If we now consider the scalars that result from the 1+1+2-decomposition, as given in Eq. (\ref{112scalars}), then we have a more complicated situation. It can, however, be understood in terms of a well studied class of cosmological models: the Locally Rotationally Symmetric (LRS) models pioneered by Ellis \cite{Ellis_1967, Stewart_1968}. A space-time is locally rotationally symmetric if it admits everywhere a one-dimensional continuous isotropy group, which essentially corresponds to rotations about a local axis of symmetry {\bf m}. The existence of such a symmetry means that all 3-vectors are proportional to {\bf m}, and all projected symmetric and trace-free 3-tensors are proportional to $m_a m_b - \frac{1}{2}M_{ab}$, where $M_{ab}$ is defined in Eq. (\ref{bigm}). In the language of the 1+1+2-decompostions described in Section \ref{subsec:112}, this means that all 2-vectors and 2-tensors must vanish. The entire dynamics of an LRS space-time is therefore described purely by the set of scalars given in Eq. (\ref{112scalars}). For this reason, we expect to recover a set of equations for our averaged scalars that can be written in a form similar to the equations that govern LRS cosmologies, with additional back-reaction terms due to the averaging of small-scale structure. We will therefore say that we are averaging to an LRS Bianchi cosmology.

The Bianchi classification contains all spatially homogeneous cosmological models (except Kantowski-Sachs) and divides them into classes labelled I to IX. The classes that admit LRS symmetry are I, II, III, V, VII and IX. Of these, the tilted LRS cosmologies can be found in classes V and VII${}_h$, and the fully isotropic cosmologies can be found in classes I, V, VII and IX. Due to their homogeneity, the only non-zero derivative of any of the scalars $S$ within these models will be a time derivatice. All the governing equations will then necessarily have the form
\begin{equation}\label{general_LRS_scalar_eq}
a_i \, \partial_t \left(N^{p_i} S_i\right) + N^{p_i + 1} \sum_j b_{ij} S_j + N^{p_i + 1}\sum_{j,k} c_{ijk} S_j  S_k = 0\,,
\end{equation}
where $S_i$ labels the allowed scalars from Eq. (\ref{112scalars}), $b_{ij}$ and $c_{ijk}$ are constants, and $a_i = 0$ or $1$ depending on whether the equation is a constraint or an evolution equation. The exponent $p_i$ of the lapse function $N$ depends on the scalar being considered: for the kinematic scalars $\left\lbrace \Theta,\Sigma,\mathcal{A},\phi,\xi\right\rbrace$ we have $p_i = 1$, and for the matter and Weyl curvature scalars, $\left\lbrace \mu, p, Q, \Pi, \mathcal{E}, \mathcal{H}\right\rbrace$ we have $p_i = 2$. These equations are a significant simplification of the full set of equations of motion, and will be the form we expect for the equations that govern the average of the scalars from our 1+1+2-decompostion. The only difference will be that we will find back-reaction terms on the right-hand side of each equation, instead of zero.

\section{Averaged equations for anisotropic cosmologies}\label{sec:averaged_equations}

In this section we will present the explicit form of the equations governing the averaged scalars from Eq. (\ref{112scalars}). These equations are valid for any hypersurface-orthogonal $\mathbf{n}$, and any choice of {\bf m}. 

\subsection{The Ricci identities for {\bf n}}

The Ricci identities taken with respect the time-like vector {\bf n} can be written as
\begin{equation*}
S_{abc} := 2\nabla_{[a}\nabla_{b]}n_c - R_{abcd}n^d = 0 \, .
\end{equation*}
Taking the trace of $S_{abc}$ over its first and third indices, and contracting with ${\bf n}$, gives the Raychaudhuri equation ($n^b S^a_{\ \ ba}=0$):
\begin{eqnarray}\label{Raychaudhuri_avg}
\hspace{-2.5cm}
&& \partial_t \left\langle N \Theta\right\rangle - \left\langle N\mathcal{A}\right\rangle\left(\left\langle N\mathcal{A}\right\rangle + \left\langle N\phi\right\rangle\right) + \frac{1}{3}\left\langle N\Theta\right\rangle^2 + \frac{3}{2}\left\langle N\Sigma\right\rangle^2 
+\frac{1}{2}\left\langle N^2\left(\mu + 3 p\right) \right\rangle - \left\langle N^2 \Lambda \right\rangle= \mathcal{B}_1 \, 
\end{eqnarray}
where $\partial_t S = N \dot{S}$ and $\dot{S} = n^a \nabla_a S$. This is an evolution equation for the expansion, written in terms of coordinate time, and is identical to the corresponding equation from LRS Bianchi cosmologies up to the back-reaction term $\mathcal{B}_1$, and the presence of the lapse function $N$, which cannot in general be set to unity. We find that the back-reaction term can be written as
\begin{eqnarray*}\label{B1}
\hspace{-2.5cm}
\mathcal{B}_1 = \frac{2}{3}{\rm Var}\left( N\Theta \right)  - \frac{3}{2}{\rm Var}&&\left(N\Sigma\right)  + {\rm Var}\left( N\mathcal{A} \right) + {\rm Cov}\left(N\mathcal{A},N\phi\right)   -2\left\langle N^2\Sigma_a\Sigma^a\right\rangle - \left\langle N^2\Sigma_{ab}\Sigma^{ab}\right\rangle \\
&& + \left\langle N^2 m^a D_a \mathcal{A}\right\rangle + \left\langle N^2 M^{ab} D_a \mathcal{A}_b \right\rangle + \left\langle N^2 \left(\mathcal{A}_a - a_a\right)\mathcal{A}^a\right\rangle + \left\langle \Theta \, \partial_t {N}\right\rangle ,
\end{eqnarray*}
which encodes all information about the influence of small-scale inhomogeneities on the acceleration of the expansion of space. This is the only equation that can be derived from $S_{abc}$ that has a counterpart in the standard approach of averaging to isotropic cosmology. As this is an equation for the acceleration of expansion, a sufficiently large (and positive) $\mathcal{B}_1$ would lead to an accelerating universe, at least within the spatial domain being considered.

By again taking the trace of $S_{abc}$ over the first and last indices, but this time contracting the middle index with respect to the space-like vector {\bf m} we can obtain a momentum constraint equation ($f_a^{\ b} S^c_{\ \ bc} m^a=0$):
\begin{eqnarray}\label{mom_constraint_avg}
&& \left\langle N^2 Q \right\rangle + \frac{3}{2}\left\langle N\phi\right\rangle\left\langle N\Sigma\right\rangle = \mathcal{B}_2 \, ,
\end{eqnarray}
where the back-reaction term is given by
\begin{eqnarray*}\label{B6}
\hspace{-2.5cm}
\mathcal{B}_2 &=& \frac{2}{3}\left\langle N^2 m^a D_a \Theta \right\rangle - \left\langle N^2 m^a D_a \Sigma \right\rangle - \frac{3}{2}{\rm Cov}\left(N\phi,N\Sigma\right)  - \left\langle N^2 M^{ab} D_a \Sigma_b\right\rangle + 2\left\langle N^2 a_b \Sigma^b\right\rangle + \left\langle N^2\Sigma_{ab}\zeta^{ab}\right\rangle. 
\end{eqnarray*}
This term describes the direct contribution from inhomogeneity to the large-scale momentum density in the direction of {\bf m}.

There are two remaining equations that can be derived from $S_{abc}$. The first of these is an evolution equation for the scalar part of the shear ($f_{\langle a}^{\ \ d} f_{b\rangle}^{\ \ e} n^c S_{dec} m^a m^b=0$):
\begin{eqnarray}\label{shear_evol_avg}
\hspace{-2.5cm}
&& \partial_t\left\langle N\Sigma\right\rangle +  \frac{2}{3}\left\langle N\Theta\right\rangle\left\langle N\Sigma\right\rangle + \frac{1}{2}\left\langle N\Sigma\right\rangle^2+\left\langle N^2\mathcal{E}\right\rangle - \frac{1}{2}\left\langle N^2\Pi\right\rangle - \frac{1}{3}\left(2\left\langle N\mathcal{A}\right\rangle - \left\langle N\phi\right\rangle \right)\left\langle N\mathcal{A}\right\rangle = \mathcal{B}_3 \, ,
\end{eqnarray}
where the back-reaction term is given by
\begin{eqnarray*}\label{B3}
\hspace{-2.5cm}
\mathcal{B}_3 = \frac{1}{3}{\rm Cov}\left(N\Theta,N\Sigma\right) + \frac{2}{3}{\rm Var}\left( N\mathcal{A} \right) - \frac{1}{3}{\rm Cov}\left(N\phi, N\mathcal{A}\right) + \frac{2}{3}\left\langle N^2 m^a D_a \mathcal{A}\right\rangle - \frac{1}{2}{\rm Var}\left(N\Sigma \right)&& \\
\nonumber  - \frac{1}{3}\left\langle N^2 M^{ab} D_a \mathcal{A}_b \right\rangle - \frac{1}{3}\left\langle N^2\Sigma_a\Sigma^a\right\rangle + \frac{1}{3}\left\langle N^2 \mathcal{A}_a\mathcal{A}^a\right\rangle + \frac{1}{3}\left\langle N^2\Sigma_{ab}\Sigma^{ab}\right\rangle  + 2\left\langle N^2\alpha_a\Sigma^a\right\rangle && \\
\nonumber  - \frac{2}{3}\left\langle N^2 a_a\mathcal{A}^a\right\rangle  + \left\langle \Sigma\, \partial_t {N} \right\rangle \, . &&
\end{eqnarray*}
This drives the generation of anisotropy in the expansion of space. The last equation to be derived from $S_{abc}$ is a constraint that is usually used to set either the acceleration of {\bf n} (i.e. $\mathcal{A}$) or the twist of {\bf m} (i.e. $\xi$) to zero. In this case, however, we find ($f_a^{\ d} f_b^{\ e} n^c S_{dec}\eta^{abf} m_f =0$):
\begin{eqnarray} \label{Axi}
&& \left\langle N\mathcal{A}\right\rangle\left\langle N\xi\right\rangle = \mathcal{B}_{4}\,,
\end{eqnarray}
where the back-reaction term is given by 
$\mathcal{B}_{4} = -\frac{1}{2}\left\langle N^2 \epsilon_{ab}D^a \mathcal{A}^b\right\rangle - {\rm Cov}\left(N\mathcal{A},N\xi\right)$,
In this case we see that any non-zero effect from the small-scale inhomogeneity will lead to a more complicated result, with $ \left\langle N \mathcal {A}\right\rangle  \neq 0 \neq \left\langle N\xi\right\rangle$.

\subsection{The Ricci identities for {\bf m}}

We can also consider the Ricci identities defined with respect to the preferred space-like direction {\bf m}:
\begin{eqnarray*}
R_{abc} := 2\nabla_{[a}\nabla_{b]}m_c - R_{abcd}m^d =0 \, .
\end{eqnarray*}
This time, the only equation that has a counterpart in isotropic cosmology is obtained by contracting over the last two indices with the screen space projection tensor, and contracting the first index with {\bf m}. This gives the Hamiltonian constraint equation ($ m^a M^{bc} R_{abc}=0 $):
\begin{eqnarray}\label{Hamiltonian_avg}
\hspace{-2.5cm}
\frac{2}{9}\left\langle N\Theta\right\rangle^2-\frac{1}{2}\left\langle N\phi\right\rangle^2+\frac{1}{3}\left\langle N\Theta\right\rangle\left\langle N\Sigma\right\rangle -\left\langle N\Sigma\right\rangle^2 - &&\frac{2}{3}\left(\left\langle N^2\mu\right\rangle + \left\langle N^2  \Lambda \right\rangle\right) \\
\nonumber && - \frac{1}{2}\left\langle N^2 \Pi\right\rangle-\left\langle N^2\mathcal{E}\right\rangle+2\left\langle N\xi\right\rangle^2 = \mathcal{B}_5 .
\end{eqnarray}
The back-reaction term in this case is given by the following expression:
\begin{eqnarray*}\label{B2}
\hspace{-2.5cm}
\mathcal{B}_5 &=& \left\langle N^2 m^a D_a \phi \right\rangle + \frac{1}{2}{\rm Var}\left( N\phi \right) - \frac{2}{9}{\rm Var}\left( N\Theta \right) + {\rm Var}\left( N\Sigma \right)- \frac{1}{3}{\rm Cov}\left(N\Theta,N\Sigma\right) - 2\,{\rm Var}\left(N\xi \right) \\
\nonumber \hspace{-2.5cm}&&\hspace{6cm}  + \left\langle N^2\zeta_{ab}\zeta^{ab}\right\rangle + \left\langle N^2 \Sigma_a\Sigma^a\right\rangle + \left\langle N^2 a_b a^b\right\rangle - \left\langle N^2 M^{ab} D_a a_b\right\rangle. 
\end{eqnarray*}
This term acts in the same way as an additional effective energy density, in what would reduce to the Friedmann equation in isotropic cosmology. Comparing (\ref{Hamiltonian_avg}) to the trace of the Gauss embedding equation, $^{(3)}R = 2\mu - \frac{2}{3}\Theta^2 + 2\sigma^2 + 2\Lambda$,
and using $\sigma^2 = \frac{3}{4}\Sigma^2 + \Sigma_a \Sigma^a + \frac{1}{2}\Sigma_{ab}\Sigma^{ab}$, one finds that the average (lapse-weighted) Ricci curvature of the space-like hypersurfaces is given by
\begin{eqnarray*}\label{meanRicci3}
\hspace{-2.5cm}
\left\langle N^2 \: ^{(3)} R \right\rangle &=& -\frac{3}{2}\left\langle N\phi\right\rangle^2 - \frac{3}{2}\left\langle N\Sigma\right\rangle^2 + \left\langle N\Theta\right\rangle\left\langle N\Sigma\right\rangle - 3\left\langle N^2\mathcal{E}\right\rangle - \frac{3}{2}\left\langle N^2\Pi\right\rangle + 6\left\langle N\xi\right\rangle^2 - 3\left\langle N^2 m^a D_a \phi\right\rangle \\
\nonumber \hspace{-2.5cm} && - \frac{3}{2}\,{\rm Var}\left( N\phi \right) - \frac{3}{2}\,{\rm Var}\left( N\Sigma \right) + {\rm Cov}\left(N\Theta,N\Sigma\right) + 6\,{\rm Var}\left( N\xi \right) - \left\langle N^2\Sigma_a\Sigma^a\right\rangle - 3\left\langle N^2 a_b a^b\right\rangle \\
\nonumber \hspace{-2.5cm} && + 3\left\langle N^2 M^{ab}D_a a_b\right\rangle - 3\left\langle N^2\zeta_{ab}\zeta^{ab}\right\rangle + \left\langle N^2 \Sigma_{ab}\Sigma^{ab}\right\rangle.
\end{eqnarray*}
The second equation to be gained from contracting $R_{abc}$ with the screen space projector is to be found using the time-like vector {\bf n}, and is an evolution for the expansion of ${\bf m}$ in the preferred space-like direction ($n^a M^{bc} R_{abc}=0$):
\begin{eqnarray}
&& \partial_t \left\langle N\phi\right\rangle - \left\langle N^2 Q\right\rangle -\left(\frac{2}{3}\left\langle N\Theta\right\rangle - \left\langle N \Sigma\right\rangle\right)\left(\left\langle N\mathcal{A}\right\rangle - \frac{1}{2}\left\langle N\phi\right\rangle\right) = \mathcal{B}_{6} \, ,
\end{eqnarray}
where we find
\begin{eqnarray*}\label{B12}
\hspace{-2.5cm}
\mathcal{B}_{6} &=& \frac{2}{3}{\rm Cov}\left(N\Theta,N\phi\right) + \frac{1}{2}{\rm Cov}\left(N\Sigma,N\phi\right) + \frac{2}{3}{\rm Cov}\left(N\Theta,N\mathcal{A}\right)-{\rm Cov}\left(N\Sigma,N\mathcal{A}\right) \\
\hspace{-2.5cm} \nonumber && + \left\langle N^2 M^{ab} D_a \Sigma_b\right\rangle  - \left\langle N^2 \zeta_{ab}\Sigma^{ab}\right\rangle + \left\langle N^2 \mathcal{A}^a\left(\alpha_a - \Sigma_a\right)\right\rangle + \left\langle N^2 a^a \left(\Sigma_a - \mathcal{A}_a\right)\right\rangle + \left\langle \phi \, \partial_t {N}\right\rangle.
\end{eqnarray*}
This equation has no counterpart in isotropic cosmological modelling.

The first of the three further equations that can be derived from $R_{abc}$ are a constraint on the magnetic part of the Weyl tensor ($\epsilon^{ab} n^c R_{abc} = 0 $):
\begin{eqnarray}\label{H_weyl_constraint}
&& \left\langle N^2 \mathcal{H}\right\rangle - 3\left\langle N\xi\right\rangle\left\langle N\Sigma\right\rangle = \mathcal{B}_{7} \, ,
\end{eqnarray}
where we find $\mathcal{B}_{7} = 3\,{\rm Cov}\left(N\xi,N\Sigma\right) + \left\langle N^2 \epsilon_{ab}\left(D^a\Sigma^b - \zeta^{ac}\Sigma^b_{\:c}\right)\right\rangle$. For our averaged cosmology to be ``silent'' we therefore require $\mathcal{B}_{7}$ to vanish along with either $\left\langle N\xi\right\rangle$ or $\left\langle N\Sigma\right\rangle$, or to conspire to cancel $3\left\langle N\xi\right\rangle\left\langle N\Sigma\right\rangle$.
The second equation is an evolution equation for the twist of {\bf m} ($n^a \epsilon^{bc}R_{abc}=0$):
\begin{eqnarray}
&& \partial_t\left\langle N\xi\right\rangle + \frac{1}{2}\left(\frac{2}{3}\left\langle N\Theta\right\rangle - \left\langle N\Sigma\right\rangle\right)\left\langle N\xi\right\rangle - \frac{1}{2}\left\langle N^2\mathcal{H}\right\rangle = \mathcal{B}_{8} \,,
\end{eqnarray}
with back-reaction term 
\begin{eqnarray*}\label{B13}
\hspace{-2.5cm}
\mathcal{B}_{8} &=& \frac{2}{3}{\rm Cov}\left( N\Theta, N\xi\right) + \frac{1}{2}{\rm Cov}\left(N\Sigma,N\xi\right)   + \frac{1}{2}\left\langle N^2 \epsilon_{ab}\left(a^a + \mathcal{A}^a\right)\left(\alpha^b+\Sigma^b\right)\right\rangle 
\\ \hspace{-2.5cm} && \hspace{6cm}  + \frac{1}{2}\left\langle N^2 \epsilon_{ab}D^a\alpha^b\right\rangle + \frac{1}{2}\left\langle N^2 \epsilon_{ac}\zeta_b^{\ c}\Sigma^{ab}\right\rangle + \left\langle \xi\,\partial_t {N}\right\rangle\,.
\end{eqnarray*}
The final equation involving $R_{abc}$ is a contraint on the product of the expansion of {\bf m} in the preferred space-like direction and its twist ($m^a \epsilon^{bc} R_{abc} = 0 $):
\begin{eqnarray}
&& \left\langle N\phi\right\rangle\left\langle N\xi\right\rangle = \mathcal{B}_{9} \, .
\end{eqnarray}
The back-reaction term in this equation is 
\begin{eqnarray*}
\mathcal{B}_{9} = -\left\langle N^2 m^a D_a \xi \right\rangle - {\rm Cov}\left(N\phi,N\xi\right) + \frac{1}{2}\left\langle N^2 \epsilon_{ab}\left(D^a a^b + \Sigma^a a^b\right)\right\rangle\,.
\end{eqnarray*}
As in Eq. (\ref{Axi}), the back-reaction term in this last equation prevents one obtaining the usual result that either $\left\langle N\phi\right\rangle$ or $\left\langle N\xi\right\rangle$ vanishes, with $\left\langle N\phi\right\rangle \neq 0 \neq \left\langle N\xi\right\rangle$ being required if $\mathcal{B}_{9}\neq 0$.

\subsection{The Bianchi identities}

Our final set of equations are derived from the Bianchi identities:
\begin{eqnarray}
W_{abcde} := \nabla_{[a}R_{bc]de} = 0 \, .
\end{eqnarray}
The only equation that could be derived from these identities in the case of isotropic cosmologies would be the energy conservation equation, which in our case takes the form ($n^a W_{abc}^{\ \ \ bc}=0$):
\begin{eqnarray}\label{Bianchi_energy_avg}
\hspace{-2.5cm}
&& \partial_t \left\langle N^2 \mu\right\rangle + \left\langle N\Theta\right\rangle\left(\left\langle N^2\mu\right\rangle+\left\langle N^2 p \right\rangle \right) + \frac{3}{2}\left\langle N\Sigma\right\rangle\left\langle N^2 \Pi\right\rangle +\left(\left\langle N\phi\right\rangle + 2\left\langle N\mathcal{A}\right\rangle\right)\left\langle N^2 Q \right\rangle = \mathcal{B}_{10} \, ,
\end{eqnarray}
where the back-reaction term is given by
\begin{eqnarray*}\label{B3}
\hspace{-2.5cm}
\mathcal{B}_{10} &=& -{\rm Cov}\left(N^2 p,N\Theta\right) - \left\langle N^3 M^{ab} D_a Q_b \right\rangle - \left\langle N^3 m^a D_a Q \right\rangle - {\rm Cov}\left(N\left(\phi+2\mathcal{A}\right),N^2Q\right)\\
\hspace{-2.5cm} \nonumber && - \frac{3}{2}{\rm Cov}\left(N\Sigma,N^2\Pi\right) - 2\left\langle N^3 \mathcal{A}_a Q^a\right\rangle - 2\left\langle N^3\Sigma_a\Pi^a\right\rangle - \left\langle N^3\Sigma_{ab}\Pi^{ab}\right\rangle \\
\hspace{-2.5cm} \nonumber && + \left\langle N^3 a_a Q^a\right\rangle + 2\left\langle N \mu\,\partial_t {N}\right\rangle\,. 
\end{eqnarray*}
The back-reaction term here provides an additional source that drives the rate of change of the average energy density, $\left\langle N^2 \mu\right\rangle$. All other equations correspond to anisotropic cosmologies only, and vanish in the isotropic limit. It can be seen that if one chooses the foliation such that $\mathbf{n}$ is the 4-velocity of pressureless dust, then $\mathcal{B}_{10}=0$. For a generic foliation, however, this need not be the case, as an observer that is not comoving with the dust will typically measure non-zero pressure, momentum density and anisotropic stress.

The first of the remaining equations gives conservation of momentum density ($m^a W_{abc}^{\ \ \ bc}=0$):
\begin{eqnarray} \label{mce}
\hspace{-2.5cm}
&& \partial_t\left\langle N^2 Q\right\rangle + \left(\frac{4}{3}\left\langle N\Theta\right\rangle + \left\langle N\Sigma\right\rangle\right) \left\langle N^2 Q\right\rangle  - \left\langle N\mathcal{A}\right\rangle\left\langle N^2\left(\mu + p + \Pi\right)\right\rangle + \frac{3}{2}\left\langle N^2 \Pi\right\rangle\left\langle N\phi\right\rangle = \mathcal{B}_{11}.
\end{eqnarray}
This equation contains the back-reaction term
\begin{eqnarray*}\label{B11}
\hspace{-2.5cm} \mathcal{B}_{11} &=& -\frac{1}{3}{\rm Cov}\left(N\Theta, N^2 Q\right) - \left\langle N^3 m^a D_a \left(p + \Pi\right) \right\rangle - \left\langle N^3 M^{ab} D_a \Pi_b\right\rangle  - {\rm Cov}\left(\frac{3}{2}N\phi+N\mathcal{A},N^2\Pi\right) \\ \hspace{-2.5cm}
\nonumber && - {\rm Cov}\left(N\Sigma, N^2 Q\right) - {\rm Cov}\left(N^2\left(\mu+p\right), N \mathcal{A}\right)  - \left\langle \mathcal N^3 {A}_a\Pi^a\right\rangle + \left\langle N^3 \zeta_{ab}\Pi^{ab}\right\rangle \\
\hspace{-2.5cm} \nonumber && + \left\langle N^3 \left(\alpha_a-\Sigma_a\right)Q^a\right\rangle + 2\left\langle N^3 a_a \Pi^a\right\rangle + 2\left\langle N Q \,\partial_t {N}\right\rangle\,. 
\end{eqnarray*}
This term drives the evolution of the momentum density, and if non-zero can therefore source a bulk flow in the averaged cosmology.

There are four remaining equations to be derived from $W_{abcde}$. Two of these are evolution equations for the scalar parts of the electric and magnetic Weyl tensor, and two are constraints for these quantities. The evolution equation for the electric part of the Weyl tensor is given by ($f^{de} n^c W_{cdabe} m^a m^b=0$):
\begin{eqnarray}\label{E_weyl_evol_avg}
&& \hspace{-2.5cm} \partial_t\left\langle N^2 \mathcal{E}\right\rangle + \frac{1}{2}\partial_t\left\langle N^2 \Pi\right\rangle + \left(\left\langle N\Theta\right\rangle - \frac{3}{2}\left\langle N\Sigma\right\rangle \right)\left\langle N^2\mathcal{E}\right\rangle + \frac{1}{2}\left(\frac{1}{3}\left\langle N\Theta\right\rangle + \frac{1}{2}\left\langle N\Sigma\right\rangle \right)\left\langle N^2\Pi\right\rangle \\
&& \hspace{-2.5cm} \nonumber - \frac{1}{3}\left(\frac{1}{2}\left\langle N\phi\right\rangle-2\left\langle N\mathcal{A}\right\rangle\right)\left\langle N^2 Q \right\rangle + \frac{1}{2}\left(\left\langle N^2 \mu\right\rangle+\left\langle N^2 p \right\rangle \right) \left\langle N\Sigma\right\rangle - 3\left\langle N\xi \right\rangle\left\langle N^2 \mathcal{H}\right\rangle = \mathcal{B}_{12} \, ,
\end{eqnarray}
with back-reaction
\begin{eqnarray*}\label{B12}
\hspace{-2.5cm}\mathcal{B}_{12} &=& -\frac{1}{3}\left\langle N^3 m^a D_a Q \right\rangle + \frac{1}{3}{\rm Cov}\left(N\Theta,N^2\Pi\right) + \left\langle N^3\epsilon_{ab}D^a \mathcal{H}^b\right\rangle + \frac{1}{6}\left\langle N^3 M^{ab} D_a Q_b\right\rangle \\
\nonumber \hspace{-2.5cm} &&  + \frac{3}{2}{\rm Cov}\left(N\Sigma,N^2\mathcal{E}\right) + \left\langle N^3\epsilon_{ab}\mathcal{H}^{bc}\zeta^a_{\ c}\right\rangle -\frac{1}{2}{\rm Cov}\left(N\Sigma,N^2\left(\mu+p+\frac{1}{2}\Pi\right)\right) \\
\nonumber \hspace{-2.5cm}&&+ \frac{1}{3}{\rm Cov}\left(N^2 Q,\frac{1}{2}N\phi-2N\mathcal{A}\right)   - \frac{1}{6}\left\langle N^3 \Sigma_a\Pi^a\right\rangle + \frac{1}{3}\left\langle N^3 \left(\mathcal{A}_a + a_a\right)Q^a\right\rangle + 2\left\langle N^3 \epsilon_{ab}\mathcal{A}^a\mathcal{H}^b\right\rangle \\
\nonumber \hspace{-2.5cm}&& - \left\langle N^3 \Sigma_{ab}\left(\mathcal{E}^{ab}+\frac{1}{2}\Pi^{ab}\right)\right\rangle  + 3\,{\rm Cov}\left(N\xi, N^2\mathcal{H}\right) + \left\langle N^3\alpha_a\Pi^a\right\rangle \\
\nonumber \hspace{-2.5cm}&&+ \left\langle N^3\left(2\alpha_a+\Sigma_a\right)\mathcal{E}^a\right\rangle + 2\left\langle N \mathcal{E}\,\partial_t N\right\rangle\,.
\end{eqnarray*}
The evolution equation for the magnetic part of the Weyl tensor is ($\eta_a^{\ cd} n^e n^f W_{ecdfb}m^a m^b=0$):
\begin{eqnarray}\label{H_weyl_evol}
\hspace{-2.5cm}
&& \partial_t\left\langle N^2\mathcal{H}\right\rangle + \left(\left\langle N\Theta\right\rangle - \frac{3}{2}\left\langle N\Sigma\right\rangle\right)\left\langle N^2\mathcal{H}\right\rangle + 3\left\langle N\xi\right\rangle\left(\left\langle N^2\mathcal{E}\right\rangle-\frac{1}{2}\left\langle N^2\Pi\right\rangle\right) = \mathcal{B}_{13} \, ,
\end{eqnarray}
which has back-reaction term
\begin{eqnarray*}\label{B13}
\hspace{-2.5cm}
\mathcal{B}_{13} &=& -\left\langle N^3\epsilon_{ab}D^a\left(\mathcal{E}^b-\frac{1}{2}\Pi^b\right)\right\rangle + \frac{3}{2}{\rm Cov}\left(N\Sigma,N^2\mathcal{H}\right) + 2\left\langle N^3\epsilon_{ab}\mathcal{E}^a \mathcal{A}^b\right\rangle + \left\langle N^3\left(2\alpha_a + \Sigma_a\right)\mathcal{H}^a\right\rangle  \\
\nonumber \hspace{-2.5cm}&& + \frac{1}{2}\left\langle N^3\epsilon_{ab}\Sigma^a Q^b\right\rangle 
- \left\langle N^3\Sigma_{ab}\mathcal{H}^{ab}\right\rangle  + \frac{1}{2}\left\langle N^3\epsilon_{ab}\mathcal{E}^{ac}\zeta^b_{\ c}\right\rangle \\
\nonumber \hspace{-2.5cm} && - 3\,{\rm Cov}\left(N\xi,N^2\mathcal{E}-\frac{1}{2}N^2\Pi\right) + 2 \left\langle N \mathcal{H} \, \partial_t {N}\right\rangle\,.
\end{eqnarray*}
The remaining two equations are a constraint equation for the electric Weyl tensor ($ f_a^{\ b} n^c n^d W_{db \ \ ec}^{ \ \ e} m^a=0 $):
\begin{eqnarray}
&& \frac{3}{2}\left(\left\langle N^2\mathcal{E}\right\rangle + \frac{1}{2}\left\langle N^2\Pi\right\rangle\right)\left\langle N\phi\right\rangle + \left(\frac{1}{3}\left\langle N\Theta\right\rangle-\frac{1}{2}\left\langle N\Sigma\right\rangle\right)\left\langle N^2 Q\right\rangle = \mathcal{B}_{14}\,,
\end{eqnarray}
which we find to have a back-reaction term given by
\begin{eqnarray*}\label{B14}
\hspace{-2.5cm}
\mathcal{B}_{14} &=& -\left\langle N^3 m^a D_a \mathcal{E}\right\rangle + \frac{1}{3}\left\langle N^3 m^a D_a \mu \right\rangle - \frac{1}{2}\left\langle N^3 m^a D_a \Pi\right\rangle - \left\langle N^3 M^{ab} D_a \left(\mathcal{E}_b+\frac{1}{2}\Pi_b\right)\right\rangle \\
\nonumber \hspace{-2.5cm}&& - \frac{3}{2}{\rm Cov}\left(N\phi,N^2\mathcal{E}+\frac{1}{2}N^2\Pi\right)+\frac{1}{2}{\rm Cov}\left(N^2 Q,N\Sigma-\frac{2}{3}N\Theta\right)+  \frac{1}{2}\left\langle N^3\Sigma_a Q^a\right\rangle \\
\nonumber \hspace{-2.5cm}&&   + \left\langle N^3 \epsilon_{ab}\Sigma^{ac}\mathcal{H}_c^{\ b}\right\rangle + \left\langle N^3\left(\mathcal{E}_{ab}+\frac{1}{2}\Pi_{ab}\right)\zeta^{ab}\right\rangle + \left\langle N^3\left(2\mathcal{E}_a+\Pi_a\right)a^a\right\rangle, 
\end{eqnarray*}
and a constraint for the magnetic part of the Weyl tensor ($\eta^{bc}_{\ \ a} n^d W_{db \ \ ec}^{\ \ e} m^a=0 $):
\begin{eqnarray}\label{hcon}
&& \frac{3}{2}\left\langle N\phi\right\rangle\left\langle N^2\mathcal{H}\right\rangle + \left\langle N^2 Q \right\rangle \left\langle N\xi\right\rangle = \mathcal{B}_{15}.
\end{eqnarray}
The final back-reaction term for this equation is given by
\begin{eqnarray*}\label{B9}
\hspace{-2.5cm}
\mathcal{B}_{15} &=& \left\langle N^3 m^a D_a \mathcal{H} \right\rangle - \frac{3}{2}{\rm Cov}\left(N\phi,N^2\mathcal{H}\right) - {\rm Cov}\left(N^2Q,N\xi\right) - \frac{1}{2}\left\langle N^3\epsilon_{ab}D^a Q^b\right\rangle \\
\nonumber \hspace{-2.5cm}&& - \left\langle N^3 M^{ab} D_a \mathcal{H}_b\right\rangle + 2\left\langle N^3 a_b \mathcal{H}^b\right\rangle + \left\langle N^3 \zeta_{ab}\mathcal{H}^{ab}\right\rangle - \left\langle N^3\epsilon_{ab}\Sigma^a_{\ c}\left(\mathcal{E}^{bc}+\frac{1}{2}\Pi^{bc}\right)\right\rangle. 
\end{eqnarray*}
Eqs. (\ref{Raychaudhuri_avg})-(\ref{hcon}) provide a complete set that can be used to describe the large-scale properties of an anisotropic cosmological model after averaging. 


These equations are a considerable complication on those that govern exact LRS cosmologies, due to the presense of the back-reaction terms, $\mathcal{B}_i$. If all of these terms are sufficiently small when calculated for a domain $\mathcal{D}$, then the expansion of $\mathcal{D}$ should be expected to be well approximated by an exact LRS Bianchi model. If this is not the case, then we will be in a situation where back-reaction of small-scale structures on the large-scale properties of the cosmology is no longer negligible. The reader may note that the equations above are all written in terms of the coordinate time $t$, which (by definition) takes the same value at every spatial averaging domain $\mathcal{D}$, as required.  In the next section we will consider a family of exact cosmological models, to help develop our understanding of these terms by explicitly calculating them for an example geometry.

\section{Example: Farnsworth dust solutions}\label{sec:farnsworth}

In order to understand our formalism, it is illustrative to apply it to a class of exact cosmological models. For this we choose the anisotropic cosmologies found by Farnsworth \cite{Farnsworth_1967}. These are exact solutions to Einstein's equations. They are of Bianchi type $V$, admitting a four-parameter group of isometries, including local rotational symmetry. The three-dimensional space-like surfaces of transitivity of this isometry group are in general not coincident with the hypersurfaces orthogonal to the dust 4-velocity: these are tilted cosmologies. The $k=-1$ FLRW metric is contained within this wider class of metrics as a special case, and although they do not contain anything that could be considered as non-linear structure, they do provide us with a precise example geometry to illustrate the application of our formalism.

The metric for the Farnworth solutions can be written as \cite{Farnsworth_1967}
\begin{equation}\label{Farnsworth_metric}
ds^2 = -dt^2 + X^2(t+Cr)dr^2 + e^{-2r}Y^2(t+Cr)\left(dy^2+dz^2\right),
\end{equation}
where the functional dependence of the metric functions $X$ and $Y$ has been fixed by the presence of the Killing vectors
\begin{equation}\label{Killing_vectors}
\hspace{-1cm}\mathbf{X}_1 = \partial_y, \quad \mathbf{X}_2 = \partial_z, \quad \mathbf{X}_3 = -C\partial_t + \partial_r + y\partial_y + z\partial_z, \quad {\rm and} \quad \mathbf{X}_4 = -z\partial_y + y\partial_z \,
\end{equation} 
and where $C$ is a constant. The 1-dimensional isotropy group can be seen to be generated by $\mathbf{X}_4$. The dust 4-velocity can be written in these coordinates as $\mathbf{u} = \partial_t$, for which the corresponding axis of rotational symmetry is given by $\tilde{\mathbf{m}} = X^{-1} \partial_r$. Substituting the metric from Eq.  (\ref{Farnsworth_metric}) into Einstein's equations, one obtains a relationship between $X$ and $Y$,
\begin{equation}\label{XY_relation}
X = k^{-1}\left(CY'-Y\right)
\end{equation}
where a prime denotes a derivative with respect to $t+Cr$ and $k$ is a positive constant, as well as the following Friedmann-like equation:
\begin{equation}\label{Friedmann}
Y'^2 = \frac{D}{3Y} + k^2,
\end{equation}
where $D$ is a positive constant related to the energy density and where we have set $\Lambda=0$.

The solution to Eq. (\ref{Friedmann}) can be written in parametric form as
\begin{eqnarray}\label{Fried_solution}
&& Y \ = \frac{Wk}{2}\left(\cosh{\eta}-1\right) \\
t + && Cr = \frac{W}{2}\left(\sinh{\eta}-\eta\right), 
\end{eqnarray}
where we have defined $W := D/3k^3$. The constant $C$ controls the size of the bulk flow, and in the limit $C \rightarrow 0$ we recover FLRW and $\eta$ becomes the usual conformal time coordinate. In what follows, when required, we will choose $C=2$ and $W=125$ in order to display numerical results.


Let us now consider re-foliating this space-time by boosting along the symmetry axis from the matter frame $\mathbf{u}$ into some new frame $\mathbf{n}$, such that
\begin{eqnarray}
n_a = \gamma\left(u_a + w_a\right) \,, \qquad {\rm where} \qquad w_a = v(t+Cr) X(t+Cr) \delta_a^{\ r} \, 
\end{eqnarray}
and $\gamma= \left(1-v^2\right)^{-1/2}$ takes its usual form in terms of the 3-velocity $v$. The functional dependence of $v$ is motivated by the symmetries of the space-time, and it can be seen that this choice of boost means that all 2-vectors and tensors vanish in the new frame. This will greatly simplify the calculation of our backreaction scalars.

After the boost, the kinematic scalars can be written as
\begin{eqnarray}
&& \hspace{-0cm} \Theta = \frac{\gamma^3}{XY}\left[Y\left\lbrace Cv'+X'+v\left(2v^2-2+Xv'-vX'\right)\right\rbrace +2\gamma^{-2}Y'\left(X+Cv\right)\right] \label{Thetav} \\
&& \hspace{-0cm} \mathcal{A} = \frac{\gamma^3}{X}\left[Xv'-v^3 X'+ v\left(X'+Cv'\right)\right] \\
&& \hspace{-0cm} \Sigma = \frac{2\gamma^3}{3XY}\left[Y\left\lbrace Cv'+X'+v\left(1+Xv'-v^2 -vX'\right)\right\rbrace - \gamma^{-2}Y'\left(X+Cv\right)\right] \\
&& \hspace{-0cm} \phi = \frac{2\gamma}{XY}\left[Y'\left(C+Xv\right) - Y\right],
\end{eqnarray}
while the matter scalars become
\begin{eqnarray}
&& \hspace{0cm} \mu = \gamma^2 \rho\left(t+Cr\right) \, ,\qquad p = \frac{1}{2}\Pi = \frac{1}{3}\gamma^2 v^2 \rho \qquad {\rm and} \qquad Q = -\gamma^2 v \rho \,, \label{Qv} 
\end{eqnarray}
where the rest mass density is given by $\rho = G_{ab} u^a u^b $. As the space-time is of Petrov type D, the electric Weyl scalar $\mathcal{E}$ is invariant under boosts, and takes the following form for all possible $v$:
\begin{equation}\label{electric_weyl_farnsworth}
\hspace{-0cm}
\mathcal{E} = \frac{X\left(Y'^2 - YY''\right)\left(C^2-X^2\right) + YY'X'\left(C^2+X^2\right) - Y^2\left(CX'+X^2 X''\right)}{3 X^3 Y^2} \,.
\end{equation} 
With these results in place, we can now consider re-foliations of the Farnsworth cosmology, and calculate the 2-scalars and their associated backreaction terms for any of them. 

\subsection{Homogeneous foliation}\label{subsec:farns_homogeneous}

Let us first consider the foliation composed of leaves that coincide with the surfaces of transitivity of the Killing vectors from Eq. (\ref{Killing_vectors}). These are spatially homogeneous 3-dimensional spaces with a time-like normal $\mathbf{n}$ that must satisfy $n_a X_3^a = 0$. This immediately implies $v = -C/X$, and results in a set of spaces of constant $\eta$. The set of non-vanishing scalars in this foliation is larger than in the foliation orthogonal to the matter flow, as it contains a bulk flow that gives rise to non-zero pressure, heat flux and anisotropic stress. The full set of scalars is $\left\lbrace \Theta, \Sigma, \mathcal{E}, \phi, \mu, p, Q,\Pi \right\rbrace$, which can be immediately obtained from Eqs. (\ref{Thetav})-(\ref{Qv}). For each of these scalars ($S$) we also know that their projected covariant derivative ($D_a S$) must vanish, due to homogeneity. This immediately implies that this foliation is both constant-density and constant-mean-curvature. 

The line-element in Eq. (\ref{Farnsworth_metric}) can be recast into coordinates adapted to this foliation by making the transformation \cite{Petrov_1964}
\begin{equation}
t = Cx + \int dT \, \beta(T) \qquad {\rm and} \qquad r = - x + \int dT \frac{\alpha(T)}{\beta(T)} \,,
\end{equation}
where $\alpha = {C}/({X^2-C^2})$ and $\beta^2 = 1 - {C^2}/({X^2-C^2})$. This results in
\begin{equation}
ds^2 = -dT^2 + A(T)dx^2 + B(T)e^{2x}\left(dy^2+dz^2\right),
\end{equation}
where $X^2 = C^2 + A$ and $Y^2 = B e^{4x} \exp{\left\{2\int dT \, {\alpha}/{\beta}\right\}}$. All quantities that were functions of $t+Cr$ are now purely functions of $T$, and moreover we have that $\mathbf{n} = \partial_T$ and $\mathbf{X}_3 = -\partial_x + y\partial_y + z\partial_z$. This makes clear the orthogonality of ${\bf n}$ with the homogeneous spaces spanned by the Killing vectors from Eq. (\ref{Killing_vectors}). The lapse function is equal to unity in these coordinates.

\begin{figure}[t!]
\centering
\includegraphics[width=0.82\linewidth]{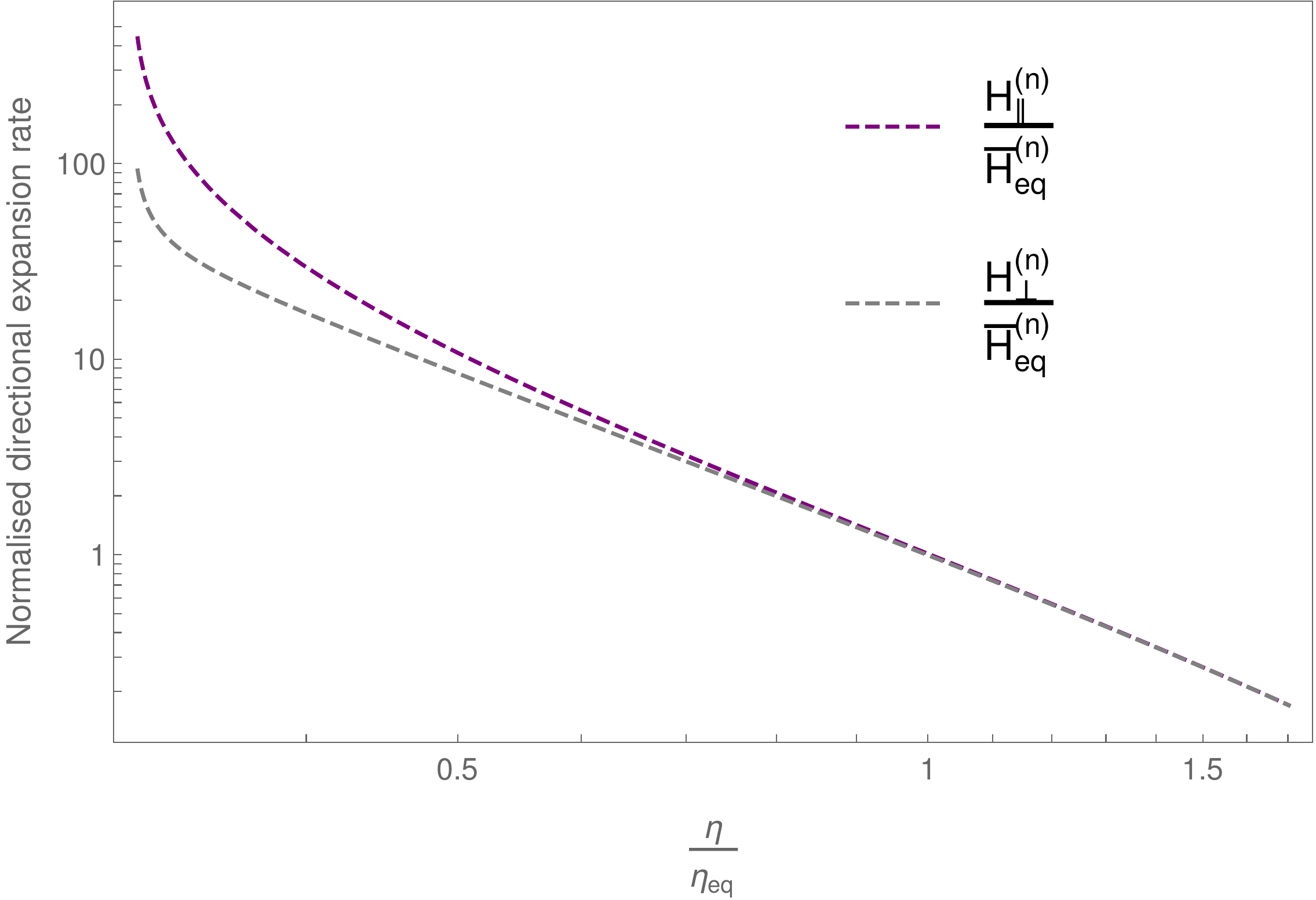}
\vspace{-0.35cm}
\caption{The normalised rate of spatial expansion in the directions parallel, $H^{(n)}_{\parallel} = \frac{1}{3}\Theta^{(n)} + \Sigma^{(n)}$, and orthogonal, $H^{(n)}_{\perp} = \frac{1}{3}\Theta^{(n)} - \frac{1}{2}\Sigma^{(n)}$, to the axis of rotational symmetry, in the homogeneous foliation of the Farnsworth solution with $\left\{C,W\right\}=\{2,125\}$.}
\label{fig:farns_directional_rates_n}
\end{figure}

\begin{figure}[h!]
\centering
\includegraphics[width=0.82\linewidth]{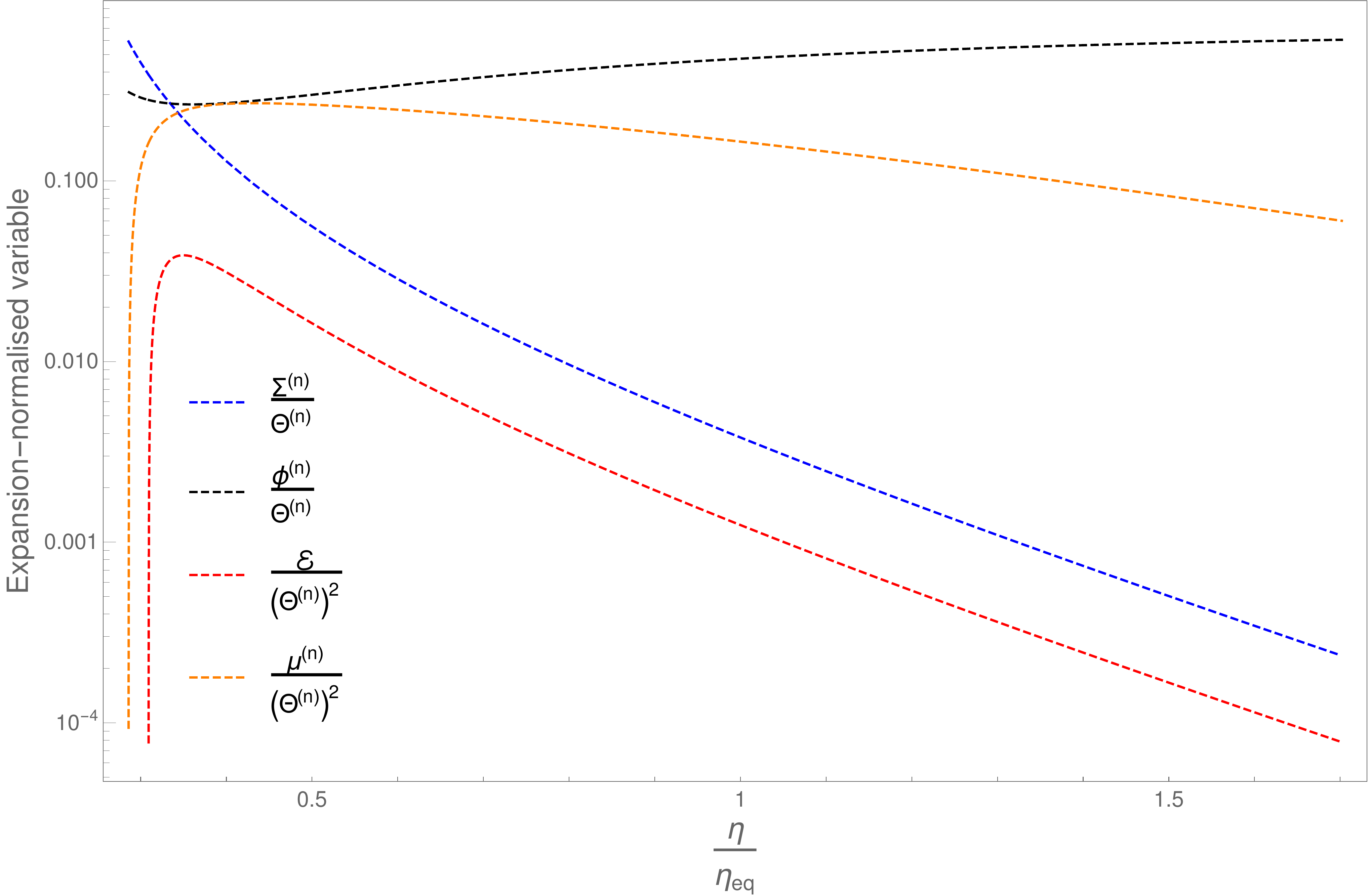}
\vspace{-0.35cm}
\caption{Expansion-normalised variables $\left\lbrace \Sigma^{(n)}, \phi^{(n)}, \mathcal{E}, \mu^{(n)} \right\rbrace$ in the homogeneous foliation of the Farnsworth solution, with $\left\{C,W\right\}=\{2,125\}$.}
\label{fig:homogeneous_functions}
\end{figure}

The behaviour of the expansion rates of space, and the expansion-normalised variables $\left\lbrace \Sigma^{(n)}, \phi^{(n)}, \mathcal{E}, \mu^{(n)} \right\rbrace$, are shown in Figures \ref{fig:farns_directional_rates_n} and \ref{fig:homogeneous_functions}. Figure \ref{fig:farns_directional_rates_n} shows the expansion rates parallel and orthogonal to the direction picked out by the bulk flow, and displays significant anisotropy at early times (small $\eta$). As the bulk flow decays at late times (large $\eta$), the difference in expansion rates in different directions also decays, demonstrating that space isotropises in this foliation. Figure \ref{fig:homogeneous_functions} shows the other kinematic scalars as a function of $\eta$, normalised by the appropriate power of the expansion rate $\Theta$ to make the resultant quantity dimensionless. One sees shear $\Sigma$ and and electric Weyl curvature $\mathcal{E}$ both increase at early times, consistent with growing anisotropy in this limit. In both plots $\eta$ is made dimensionless by normalising it relative to $\eta_{\rm eq}$, its value when $Y=-D/3k^2$ (such that the two terms on the right-hand side of Eq. (\ref{Friedmann}) are equal). In Figure \ref{fig:farns_directional_rates_n} the expansion rates are made dimensionless by normalising them with respect to the mean expansion rate $\bar{H}^{(n)} =\frac{1}{3} \Theta^{(n)}$ at the equality time $\eta_{\rm eq}$. 

In this foliation it can be seen that all of back-reaction scalars $\mathcal{B}_i$ from Section \ref{subsec:LRS} must be equal to zero. This happens because all variances and covariances must vanish (as $\int_{\mathcal{D}} d^3 x\,\sqrt{f}\, S(T) = S(T) V_{\mathcal{D}}$), and because all terms containing projected covariant derivatives (such as $m^a D_a S$) must also vanish. The cosmology one observes in this foliation can therefore be entirely described by the locally defined scalar quanitites, without any need for averaging. This is as exactly as expected for a space-time that admits a homogeneous foliation. Of course, in most geometries such a simplifying set of symmetries will not exist, and in more realistic cosmologies one would need to perform explicit averages in order to gain a set of quantities that could be used to describe the large-scale properties of space. Let us now consider an inhomogeneous foliation of the Farnsworth solutions, to show how this would work in this simple example space-time.

\subsection{Matter-rest-space foliation}\label{subsec:farns_matter}

\begin{figure}[t!]
\centering
\includegraphics[width=0.85\linewidth]{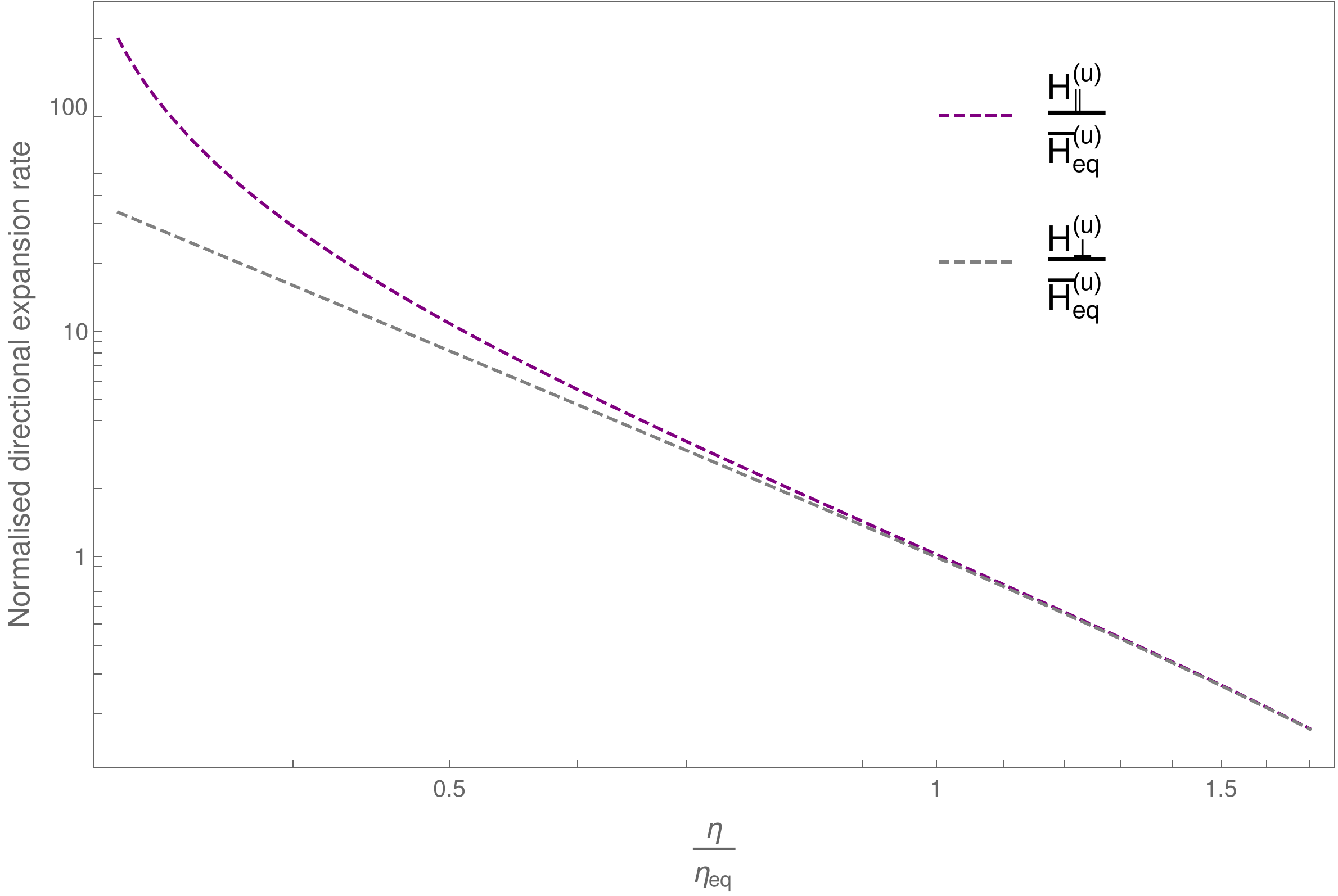}
\vspace{-0.3cm}
\caption{The normalised rate of spatial expansion in the directions parallel, $H^{(u)}_{\parallel} = \frac{1}{3}\Theta^{(u)} + \Sigma^{(u)}$, and orthogonal, $H^{(u)}_{\perp} = \frac{1}{3}\Theta^{(u)} - \frac{1}{2}\Sigma^{(u)}$, to the axis of rotational symmetry, in the matter-rest-space foliation of the Farnsworth solution with $\left\{C,W\right\}=\{2,125\}$.}
\label{fig:farns_directional_rates_u}
\vspace{-0.3cm}
\end{figure}

\begin{figure}[t!]
\centering
\includegraphics[width=0.85\linewidth]{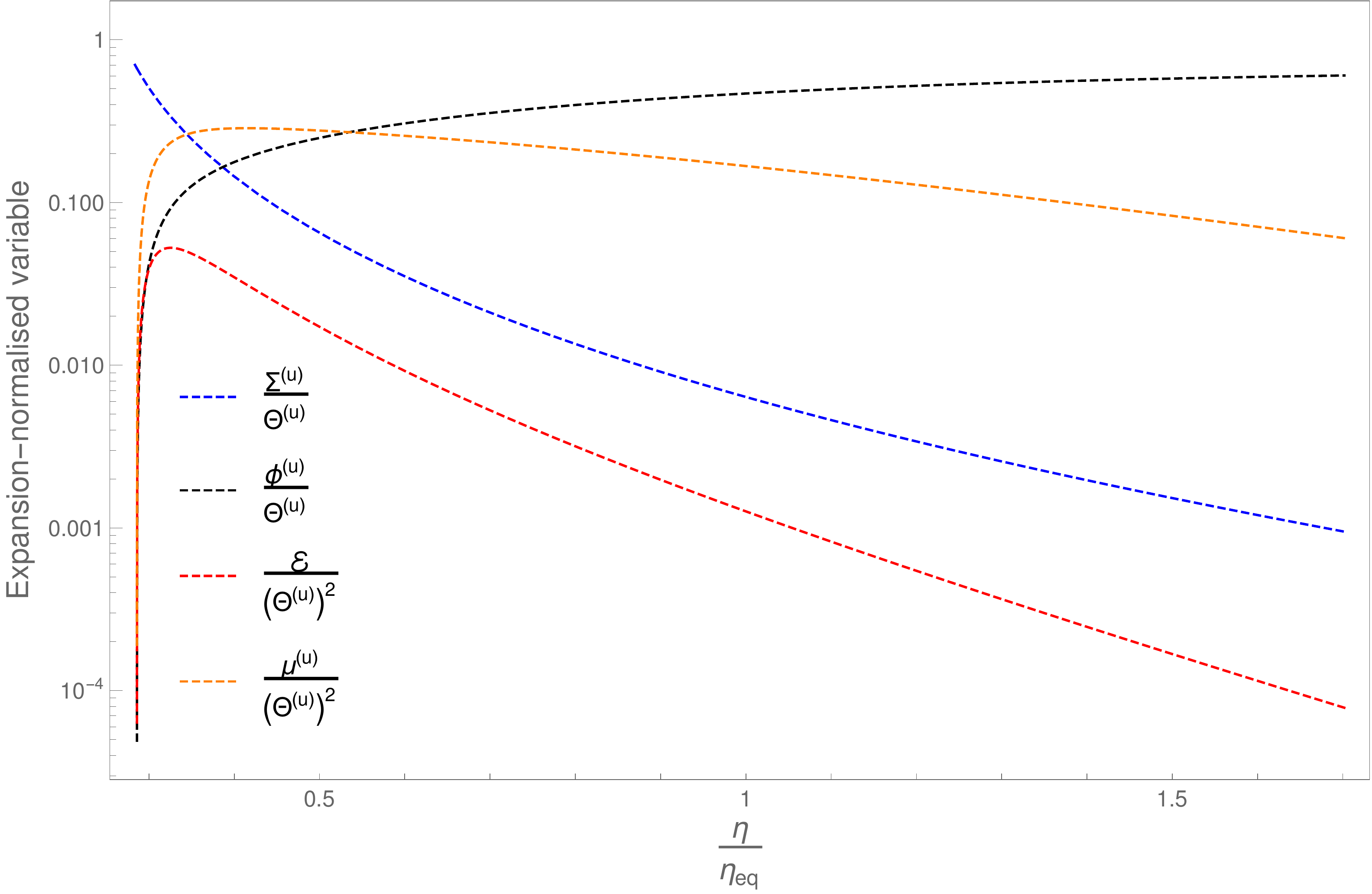}
\vspace{-0.3cm}
\caption{Expansion-normalised variables $\left\lbrace \Sigma^{(u)},\phi^{(u)},\mathcal{E},\mu^{(u)} \right\rbrace$ in the matter-rest-space foliation of the Farnsworth solution, with $\left\{C,W\right\}=\{2,125\}$.}
\label{fig:farns_functions_eta}
\vspace{-0.3cm}
\end{figure}

Let us now foliate the exact same space-time into hypersurfaces orthogonal to the matter flow ${\bf u}$, with induced metric $h_{ab}=g_{ab} + u_au_b$. This corresponds to $v=0$ in Eqs. (\ref{Thetav})-(\ref{Qv}), and to surfaces of constant coordinate time $t$. The lapse is again equal to unity, so covariant and coordinate time derivatives are equivalent. The only non-zero quantities in this case are the scalars $\left\lbrace \Theta, \Sigma, \mathcal{E}, \mu, \phi \right\rbrace$, which in terms of the parameters from the solution in Eq. (\ref{Fried_solution}) are
\begin{eqnarray}
\Theta &=& \frac{2\left[3W\sinh{\eta}\left(\cosh{\eta}-1\right)+2C\left(2\cosh{\eta}+1\right)\right]}{W\left(\cosh{\eta}-1\right)\left[W\left(\cosh{\eta}-1\right)^2-2C\sinh{\eta}\right]} \\
\Sigma &=& \frac{8C\left(\sinh^2{\eta}+\cosh{\eta}-1\right)}{3W\left(\cosh{\eta}-1\right)^2\left[W\left(\cosh{\eta}-1\right)^2-2C\sinh{\eta}\right]} \\
\phi &=& \frac{4}{W\left(\cosh{\eta}-1\right)},
\end{eqnarray}
and the dust density 
\begin{equation}\label{rho_Farnsworth}
\mu = \rho = \frac{24}{W\left(\cosh{\eta}-1\right)\left[W\left(\cosh{\eta}-1\right)^2-2C\sinh{\eta}\right]}.
\end{equation}
The electric Weyl curvature is given by Eq. (\ref{electric_weyl_farnsworth}). The normalised expansion rates as a function of $\eta$ are shown in Figure \ref{fig:farns_directional_rates_u}, and the other expansion-normalised scalars are shown in Figure \ref{fig:farns_functions_eta}. These plots, displaying quantities calculated in the foliation generated by the matter flow ${\bf u}$, may be directly compared with Figures \ref{fig:farns_directional_rates_n} and \ref{fig:homogeneous_functions}.

\begin{figure}[t!]
\centering
\includegraphics[width=0.8\textwidth]{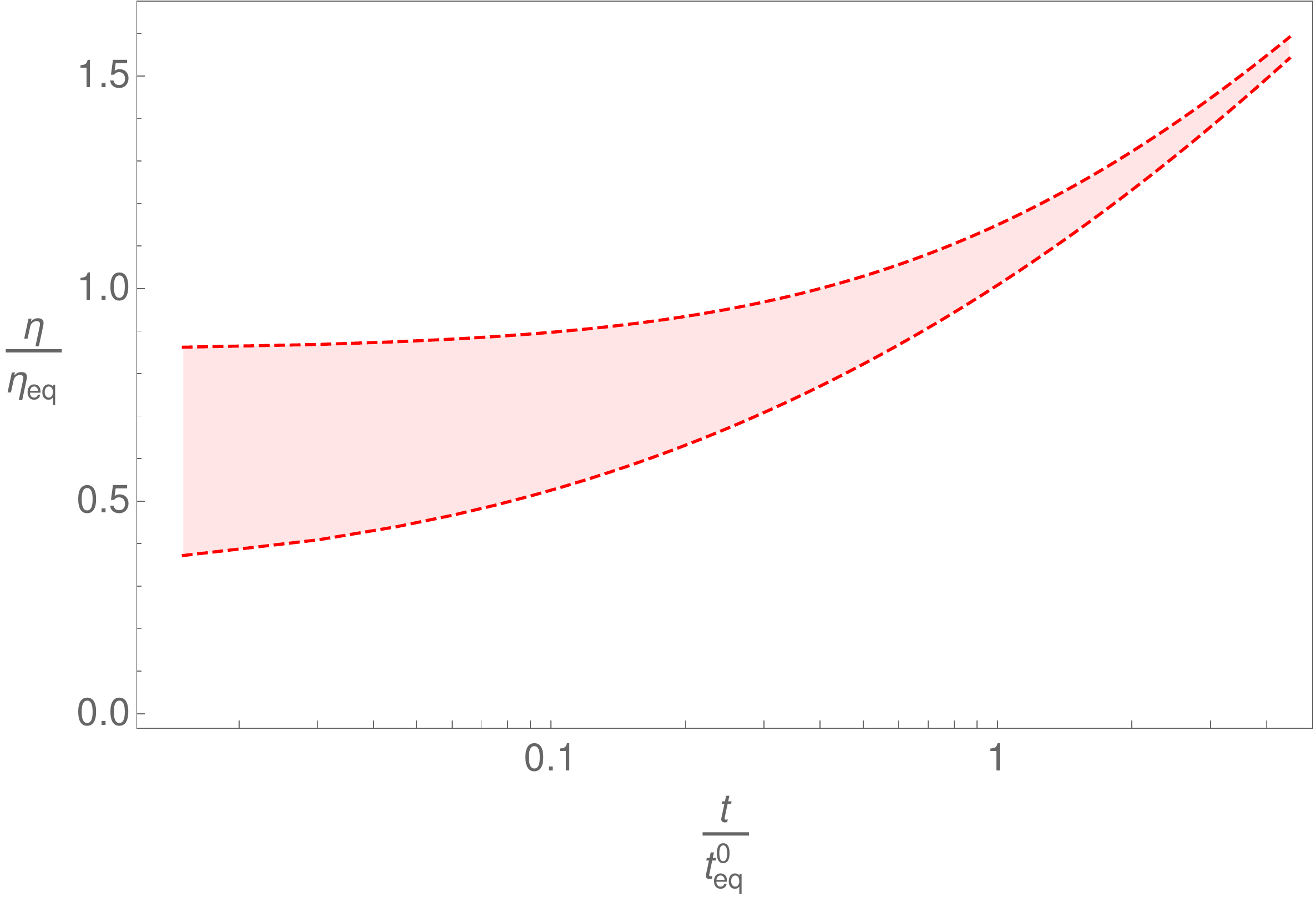}
\vspace{-0cm}
\caption{The integration domain from $r_{\rm min}=1$ to $r_{\rm max}=20$ (shaded region) in the matter-rest-frame foliation as a function of time $t$, normalised relative to $t^0_{\rm eq}=W \left(\sinh{\eta_{\rm eq}}-\eta_{\rm eq}\right)/2$, and with $\left\{C,W\right\}=\{2,125\}$.}
\label{fig:integrationlimits}
\end{figure}

In order to average these quantities we need an averaging domain $\mathcal{D}$ on each of the surfaces orthogonal to $\mathbf{u}$. If we take this domain to have coordinate extents $\left\lbrace \Delta y, \Delta z, \Delta r\right\rbrace$, then the spatial volume is given by
\begin{eqnarray}
V_{\mathcal{D}}(t) &=& \int_{\mathcal{D}} \mathrm{d}y\mathrm{d}z\mathrm{d}r \sqrt{h(\eta)} = \frac{\Delta y \Delta z W^3 k^2}{16 C} \,e^{2t/C} \, \int_{\eta(r_{\rm min},t)}^{\eta(r_{\rm max}, t)} d\eta \, J(\eta),
\end{eqnarray}
where $J(\eta) = \left(\cosh{\eta}-1\right)^2 \left[W\left(\cosh{\eta}-1\right)^2 - 2C\sinh{\eta}\right]\exp{\left\{-{W}\left(\sinh{\eta} - \eta\right)/C\right\}}$.
We can then compute all the desired scalar averages as \footnote{In the isotropic case $C \longrightarrow 0$, the integration limits $\eta(r_{\rm min},t)$ and $\eta(r_{\rm max}, t)$ coincide. However, in that case $S(\eta) \rightarrow S(t)$ can be brought outside of the integral, such that one safely obtains $\left\langle S(t)\right\rangle = S$.}
\begin{equation}
\left\langle S(t)\right\rangle = \frac{\int_{\eta(r_{\rm min},t)}^{\eta(r_{\rm max}, t)} d\eta \, J(\eta) S(\eta)}{\int_{\eta(r_{\rm min},t)}^{\eta(r_{\rm max}, t)} d\eta \, J(\eta)}\, ,
\end{equation}
where $\Delta r = r_{\rm max} - r_{\rm min}$. In choosing the integration domain (i.e. $r_{\rm min / max}$), it is important to avoid the big bang singularity that is located at $r_{\rm BB}(t)$ such that $t + C r_{\rm BB}(t) = 0$. For any $C>0$ this is satisfied by $r_{\rm min} >0$. The value of $r_{\rm max}$ needs only to be sufficiently large that $\eta_{\rm min}$ and $\eta_{\rm max}$ are relatively far apart, but once this is ensured to be true the backreaction effects are only weakly dependent on its value. If we choose $r_{\rm min} = 1$, $r_{\rm max} = 20$ and $C=2$ as an example, then one obtains the integration domain shown in Figure \ref{fig:integrationlimits}. At early times the distance from $r_{\rm min}$ to $r_{\rm max}$ corresponds to a large separation in $\eta$, while at late times the integration domain shrinks (as the space isotropises).

\begin{figure}[t!]
\centering
\includegraphics[width=0.82\textwidth]{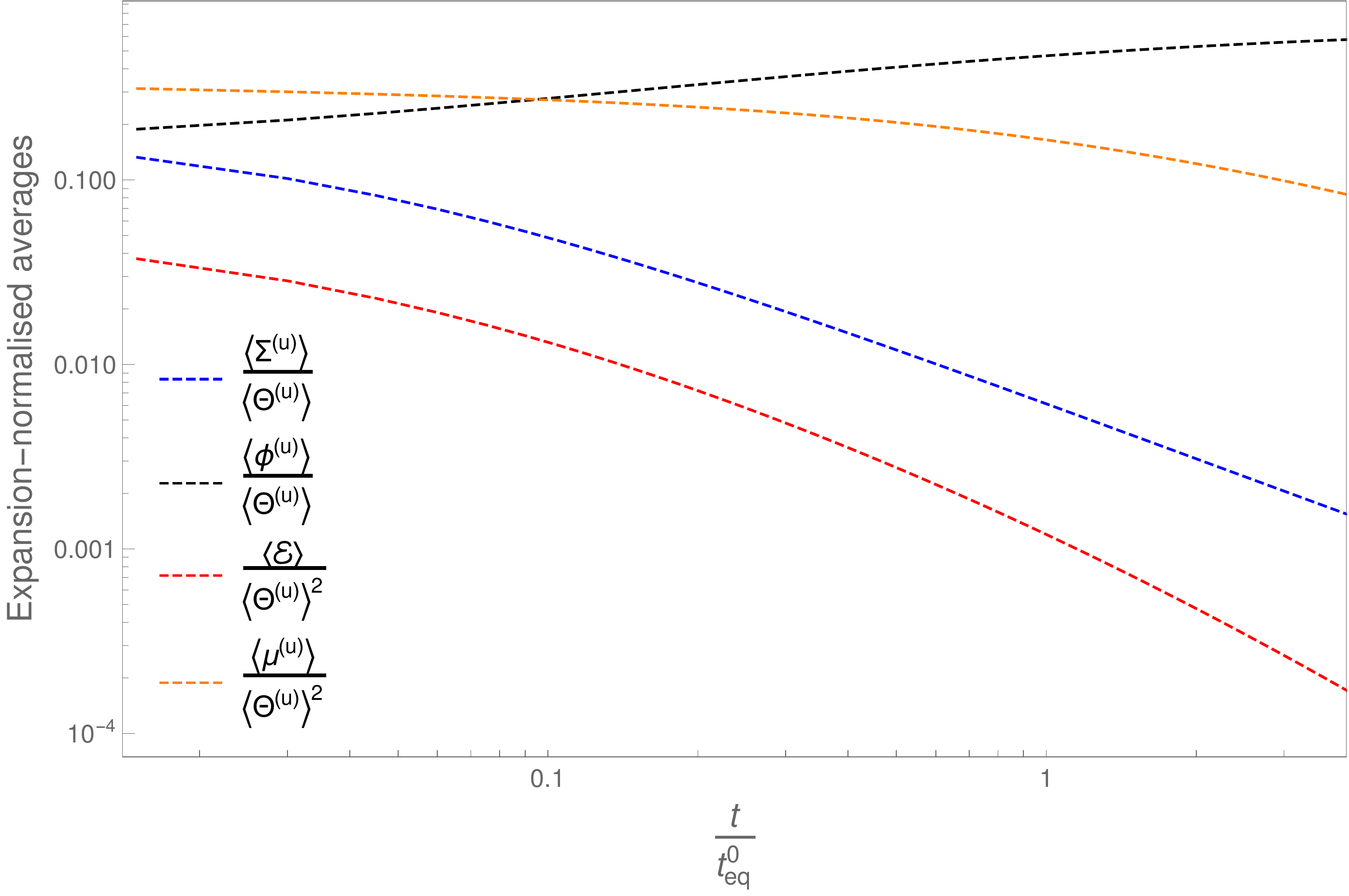}
\vspace{-0.3cm}
\caption{Evolution of $\left\lbrace \left\langle \phi^{(u)}\right\rangle,\left\langle \Sigma^{(u)}\right\rangle,\left\langle \mathcal{E}\right\rangle,\left\langle \mu^{(u)}\right\rangle \right\rbrace$ with $t$, normalised by the relevant power of $\left\langle \Theta^{(u)}\right\rangle$. The time coordinate $t$ is normalised by $t^0_{\rm eq}$. The averaging domain is from Figure \ref{fig:integrationlimits}.}
\label{fig:farns_means_normalised}
\end{figure}

\begin{figure}[t!]
\centering
\includegraphics[width=0.82\textwidth]{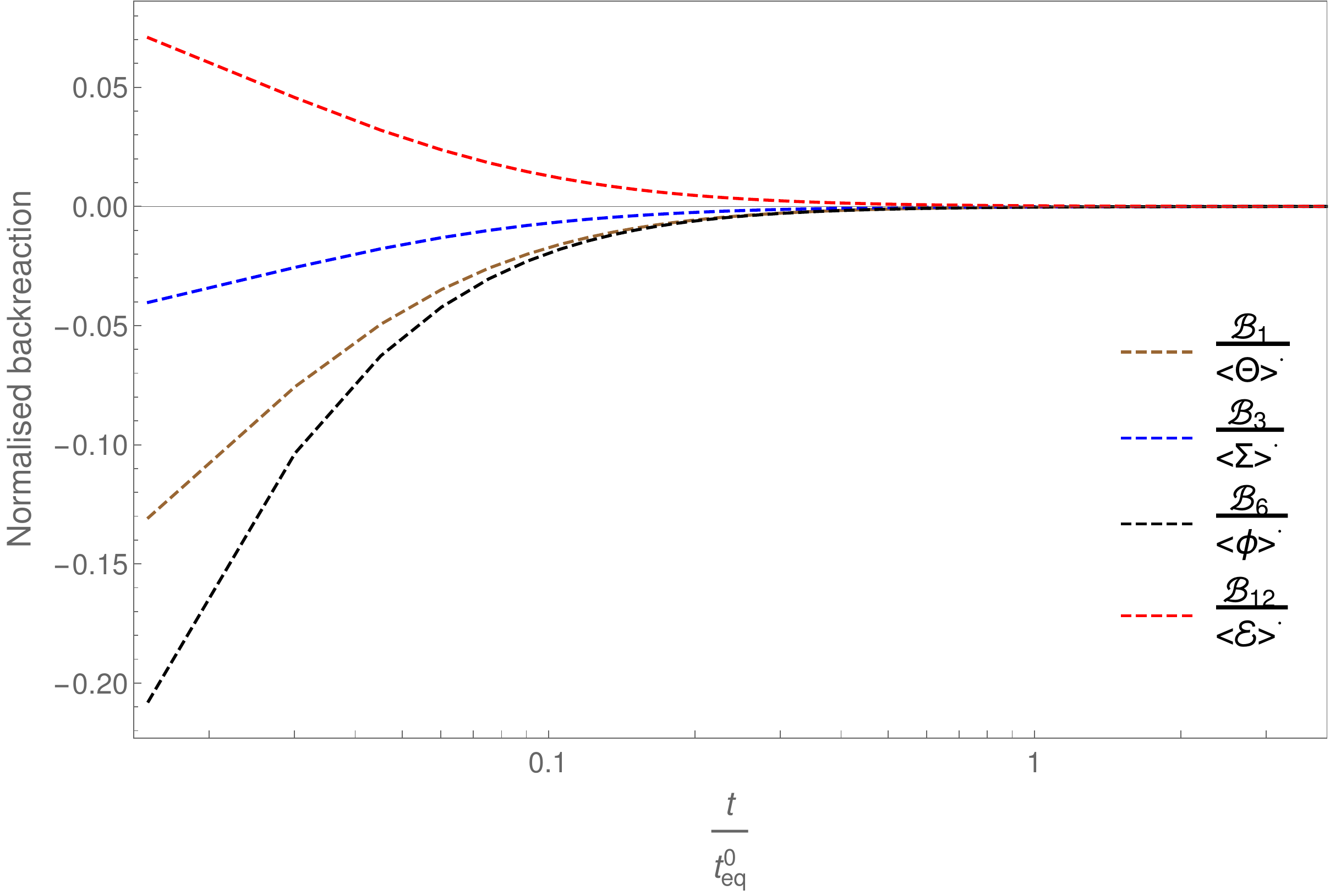}
\vspace{-0.3cm}
\caption{Back-reaction scalars $\left\lbrace \mathcal{B}_1,\mathcal{B}_3,\mathcal{B}_6,\mathcal{B}_{12}\right\rbrace$ as a function of $t/t^0_{\rm eq}$, normalised by the time derivatives of the averaged quantities whose evolution equation they appear in.The averaging domain is from Figure \ref{fig:integrationlimits}, and we have used $\left\{C,W\right\}=\{2,125\}$.}
\label{fig:farns_b1b3b6b12}
\end{figure}

The evolution with $t$ of our averaged scalars, normalised by the averaged expansion $\left\langle\Theta\right\rangle$, is shown in Figure \ref{fig:farns_means_normalised}. The inhomogeneity in these spaces has been averaged out to give a set of purely time-dependent scalar functions. These averages provide a cosmological description of the large-scale properties of the spaces orthogonal to the matter flow {\bf u}, including anisotropy. The magnitude of the back-reaction scalars that produce them are displayed in Figure \ref{fig:farns_b1b3b6b12}, where we focus on $\left\lbrace\mathcal{B}_1,\mathcal{B}_3,\mathcal{B}_6,\mathcal{B}_{12}\right\rbrace$ as they appear in evolution rather than constraint equations. These quantities are shown as ratios of the time derivative of the relevant average, as per the equations in Section \ref{sec:averaged_equations}, in order to show the fractional contribution they make to their evolution. In each case, the averaging domain used is the one displayed in Figure \ref{fig:integrationlimits}, and we have again used $\left\{C,W\right\}=\{2,125\}$.

It can be seen from Figure \ref{fig:farns_b1b3b6b12} that the back-reaction effect becomes large at early times, and dissipates to zero at late times. This is as expected in the Farnsworth solution, as anisotropy (and hence inhomogeneity in the matter-rest-space foliation) is entirely due to the bulk flow $v=-C/X$, which becomes small as the scale factor $X$ grows and the space isotropises. This example demonstrates that the large-scale averaged properties of a space depend on the chosen foliation, and that one requires in addition a choice of averaging domain (and in general a spatial direction {\bf m}). The kinematic and matter quantities that result can display different behaviours when calculated on different foliations, and so can the magnitude of back-reaction scalars. This clearly demonstrates the need for these choices to be made carefully, and for the interpretation of scalar averaging results in cosmology to be understood in terms of these choices.

\section{Discussion}\label{sec:discussion}

We have introduced an approach for modelling the emergence of cosmic anisotropy from inhomogeneous space-times. This has been achieved by performing a $1+1+2$-decomposition of all relevant fields \cite{Clarkson_2007}, and averaging the covariantly defined scalars that result \cite{Buchert_2000}. The equations that result describe a locally rotationally symmetric Bianchi cosmological model, with additional source terms due to the back-reaction from inhomogeneity. Our approach is very general, allowing for averaging surfaces and the direction of anisotropy to be freely chosen, and does not require any special properties from the matter content of the space-time.One can freely convert between different choices of foliation, according to the transformation rules set out in Section \ref{subsec:tilted_averaging}, in order to accommodate a variety of different types of cosmological model. A set of models of particular interest will be near-FLRW geometries with small tilt (i.e. with $v \ll c$), in which case one may expect our general expressions to simplify considerably.

We expect our formalism to be of use for modelling situations in which astronomical observations suggest large-scale cosmological anisotropy \cite{Aluri_2022}, as well as for assessing the significance of potential inconsistencies between different data sets \cite{Hubble1, Hubble2, Hubble}. In the former case, the existence of large-scale anisotropy has been claimed by some to be enough to nullify at least some of the evidence for dark energy \cite{Colin_2019, Mohayaee_2020, Mohayaee_2021}, while in the latter case the Hubble tension points to a possible failure of a single FLRW model to consistently account for all cosmological observables \cite{Wagner:2022etu, Krishnan:2021dyb}. Our formalism provides a way to model deviations from FLRW that may be of interest for understanding these observational anomalies, within a fully relativistic and covariant framework, and without introducing any exotic new physics. It remains to be seen whether any particular model will be able to produce an emergent anisotropy that would be compatible with any of the anomalies claimed in the literature, or whether any of the additional degrees of freedom in this approach would be sufficient to alleviate any tensions. Our framework provides a mechanism by which such questions can be investigated.


The set of equations presented in Section \ref{sec:averaged_equations} for our emergent scalars is, in general, not closed, although there do exist integrability conditions that relate the backreaction scalars $\mathcal{B}_i$ to one another. This non-closure is to be expected on mathematical grounds, as they do not contain evolution or constraint equations for the 2-vectors or tensors that result from the $1+1+2$-decomposition. On more physical grounds, one would not expect equations that govern the large-scale average of a wide class of inhomogeneous equations to be closed, as this would suggest that all inhomogeneous space-times would have the same average properties, which cannot possibly be true. In our formalism, the set of large-scale cosmological properties is dependent on the presence of non-linear structure through the $\mathcal{B}_i$. Once these are specified, the averaged cosmological equations become closed, and can be integrated.

The magnitude of back-reaction terms will depend on the geometry of the underlying spacetime. In the case of nearly FLRW cosmologies, which have been extensively studied, most authors find the relative size of back-reaction terms to be small in the Buchert equations of isotropic cosmology, such that $\mathcal{B}/H^2 \sim 10^{-5}$ (see e.g. \cite{Adamek_2019}). While such contributions are indeed small, they are not as small as one may have naively  assumed, given that $\mathcal{B}$ itself has leading-order contributions at second-order in cosmological perturbation theory. The reason why $\mathcal{B}$ is not even smaller is that it contains terms $\sim \Phi \, D^2 \Phi \sim \Phi \, \delta$, where $\delta$ is a density constrast that can become of order unity (or larger) in the presence of non-linear structures. In the present case we have back-reaction terms appearing not only in the averaged versions of the Friedmann equations, but also (for example) in the momentum evolution equation (\ref{mce}). Using the same logic, the presence of non-linear structure might be expected to result in back-reaction terms of order $\mathcal{B}_{11}/H^2 \sim 10^{-5}$. In this case, however, such a contribution would be of the same size as the usual terms on the left-hand side of Eq. (\ref{mce}), meaning that the back-reaction terms could potentially be more significant overall. Whether these expectations are, in fact, realised will require a detailed study.

In Section \ref{sec:farnsworth} we applied our formalism to a simple exact cosmological solution: the Farnsworth solution \cite{Farnsworth_1967}. This explicitly demonstrated the dependence that arises from choice of foliation and averaging domain, as well as how we envisage the formalism to be applied. It shows the necessity of choosing space-like surfaces carefully when considering concepts such as the expansion of space and bulk flows in non-FLRW cosmological models, and the effect that back-reaction from inhomogeneity can have on these quantities. In future work, we will consider more general inhomogeneous cosmological models, as well as how our formalism should be applied in the real Universe. This will necessarily require developing an understanding of how to interpret observations in terms of averages, and how to most appropriately choose both the foliation and the preferred spatial direction.

\section*{Acknowledgments}\label{sec:acknowledgments}

Calculations in this paper were performed using the tensor algebra package \texttt{diffgeo} \cite{diffgeo}. We thank Chris Clarkson for helpful discussions, and acknowledge support from STFC grant ST/P000592/1.

\newpage
\appendix

\section{Summary of notation} \label{sec:notation}

This appendix provides an exhaustive list of all covariantly defined quantities used in the text.

\subsection{Basic objects for the 1+1+2 decomposition}

\begin{itemize}

\item Preferred (hypersurface-forming) time-like vector: $\mathbf{n}$.

\item Preferred space-like vector: $\mathbf{m}$.

\item Projection tensor into space-like hypersurfaces orthogonal to $\mathbf{n}$: $f_{ab} = g_{ab} + n_a n_b$.

\item Projected covariant derivative on space-like hypersurfaces: $D_a T_{b c} = f_a^{\phantom{a} d}  f_b^{\phantom{b} e}  f_c^{\phantom{c} f} \nabla_d T_{ef}$.

\item Proper time derivative along $\mathbf{n}$: $\dot{\phantom{a}} = n^a \nabla_a$.

\item Projection tensor into screen spaces orthogonal to both $\mathbf{n}$ and $\mathbf{m}$: $M_{ab} = f_{ab} - m_a m_b$.

\item Alternating tensor on space-like hypersurfaces: $\eta_{abc} = n^d \eta_{dabc}$.

\item Alternating tensor on screen space: $\epsilon_{ab} = \eta_{abc}m^c$.

\end{itemize}

\subsection{Kinematic variables for time-like vector}

\begin{itemize}

\item Expansion of space-like hypersurfaces: $\Theta = D_a n^a$.

\item Volume-preserving shear: $\sigma_{ab} = D_{\langle a}n_{b\rangle} = \left(f_{(a}^{\ c} f_{b)}^{\ d} - \frac{1}{3}f_{ab}f^{cd}\right)\nabla_c n_d$.

\item Scalar part of shear: $\Sigma = \sigma_{ab} m^a m^b$.

\item Vector part of shear: $\Sigma_a = M_{ab} m_c \sigma^{bc}$.

\item Tensor part of shear: $\Sigma_{ab} = \left(M^{\ c}_{(a}M^{\ d}_{b)} - \frac{1}{2}M_{ab}M^{cd}\right)\sigma_{cd}$.

\item Vorticity: $\omega_a = \frac{1}{2}\eta_{abc} D^b n^c$ (choosen to be zero throughout, for {\bf n}). 

\item Scalar part of vorticity: $\Omega = \omega_a m^a = 0$.

\item Vector part of vorticity: $\Omega_a = M_{ab} \omega^b = 0$. 

\item Acceleration: $\dot{n}_a = n^b \nabla_b n_a$.

\item Scalar part of acceleration: $\mathcal{A} = \dot{n}_a m^a$.

\item Vector part of acceleration: $\mathcal{A}_a = M_{ab}\dot{n}^b$.

\end{itemize}

\subsection{Kinematic variables for space-like vector}

\begin{itemize}

\item Expansion of screen space along $\mathbf{m}$: $\phi = M^{ab} D_a m_b$.

\item Area-preserving shear of screen space along $\mathbf{m}$: $\zeta_{ab} = M^{\ c}_{(a}M^{\ d}_{b)} D_c m_d - \frac{1}{2}\phi M_{ab}$.

\item Twist of screen space: $\xi = \frac{1}{2}\epsilon_{ab} M^{ac} M^{bd} D_c m_d$.

\item Non-geodesy of $\mathbf{m}$: $a_b = m^c D_c m_b$.


\item Scalar part of $\dot{m}_a$: $\mathcal{A} = -n_a\dot{m}^a$, since $n_a m^a = 0$.

\item Vector part of $\dot{m}_a$: $\alpha_a = M_a^{\ b} \dot{m}_b$.

\end{itemize}

\subsection{Changing foliation}

\begin{itemize}

\item New time-like vector (not necessarily hypersurface-forming): $\mathbf{u}$.

\item Projection tensor orthogonal to $\mathbf{u}$: $h_{ab} = g_{ab} + u_a u_b$. 

\item Relation between time-like vectors: $n_a = \gamma\left(u_a + w_a\right) \ \Leftrightarrow \ u_a = \gamma\left(n_a - v_a\right)$, where $\gamma = \left(1 - v_a v^a\right)^{-1/2} = \left(1 - w_a w^a \right)^{-1/2}$. 

\item Boost vectors relating $\mathbf{n}$ and $\mathbf{u}$: $w_a = \gamma^{-1} h_a^{\ b} v_b$, and $v_a = \gamma^{-1} f_a^{\ b} w_b$.

\item Relation between space-like vectors, projected into surfaces orthogonal to the time-like vectors: $\tilde{m}_a = k h_a^{\ b} m_b$, and $m_a = \tilde{k}f_a^{\ b} \tilde{m}_b$, where $k = \left[1+\left(\gamma m_c v^c\right)^2\right]^{-1/2}$ and $\tilde{k} = \left[1+\left(\gamma \tilde{m}_c w^c\right)^2\right]^{-1/2}$.

\item Alternating tensor on two-dimensional surfaces orthogonal to $\mathbf{n}$ and $\mathbf{v}$: $\chi_{ab} = \eta_{abc} v^c$.

\item Expansion of boost $\mathbf{v}$ (i.e. velocity divergence): $\kappa = D_a v^a$.

\item Shear of $\mathbf{v}$: $\beta_{ab} = D_{\langle a}v_{b \rangle}$.

\item Vorticity of $\mathbf{v}$: $W_a = \frac{1}{2}\eta_{abc} D^{a}v^{b}$.


\end{itemize}

\subsection{Weyl curvature variables}

\begin{itemize}

\item Electric part of Weyl tensor: $E_{ab} = C_{acbd} n^c n^d$.

\item Scalar part of $E_{ab}$: $\mathcal{E} = E_{ab} m^a m^b$.

\item Vector part of $E_{ab}$: $\mathcal{E}_a = M_{ab} m_c E^{bc}$.

\item Tensor part of $E_{ab}$: $\mathcal{E}_{ab} = \left(M^{\  c}_{(a}M^{\ d}_{b)} - \frac{1}{2}M_{ab}M^{cd}\right)E_{cd}$.

\item Magnetic part of Weyl tensor: $H_{ab} = \frac{1}{2}\eta_{efda}C^{ef}_{\quad bc} n^c n^d$.

\item Scalar part of $H_{ab}$: $\mathcal{H} = H_{ab} m^a m^b$.

\item Vector part of $H_{ab}$: $\mathcal{H}_a = M_{ab} m_c H^{bc}$.

\item Tensor part of $H_{ab}$: $\mathcal{H}_{ab} = \left(M^{\  c}_{(a}M^{\ d}_{b)} - \frac{1}{2}M_{ab}M^{cd}\right)H_{cd}$.

\end{itemize}

\subsection{Matter variables}

\begin{itemize}

\item Energy density: $\mu = T_{ab} n^a n^b$. 

\item Isotropic pressure: $p = \frac{1}{3} T_{ab} f^{ab}$.

\item Heat flux/momentum density: $q_a = - T_{bc} n^b f_a^{\ c}$.

\item Scalar part of heat flux: $Q = q_a m^a$.

\item Vector part of heat flux: $q_a = M_{ab} q^b$.

\item Anisotropic stress: $\pi_{ab} = T_{\langle ab \rangle} = \left(f_{(a}^{\ c} f_{b)}^{\ d} - \frac{1}{3}f_{ab}f^{cd}\right)T_{cd}$.

\item Scalar part of anisotropic stress: $\Pi = \pi_{ab} m^a m^b$.

\item Vector part of anisotropic stress: $\Pi_a = M_{ab} m_c \pi^{bc}$.

\item Tensor part of anisotropic stress: $\Pi_{ab} = \left(M^{\  c}_{(a}M^{\ d}_{b)} - \frac{1}{2}M_{ab}M^{cd}\right)\pi_{cd}$.

\end{itemize}

\section*{References}

\bibliographystyle{unsrt}

\end{document}